\documentclass[superscriptaddress,showpacs,pra,onecolumn]{revtex4}
\usepackage{amsmath}
\usepackage{amssymb}
\usepackage{amsfonts}
\usepackage{amscd}
\usepackage{latexsym}
\usepackage{revsymb}
\usepackage{graphicx}

\newtheorem{defi}{Definition}
\newtheorem{lem}[defi]{Lemma}
\newtheorem{thm}[defi]{Theorem}
\newenvironment{pf}{\noindent{\bf Proof}\quad}{\hfill$\Box$\bigskip}

\input{tcilatex}

\begin{document}

\title{Universal distortion-free entanglement concentration}
\author{ Keiji Matsumoto}
\affiliation{Quantum Computation Group, National Institute of Informatics, Tokyo, Japan}
\affiliation{Quantum Computation and Information Project, ERATO, JST, 5-28-3 Hongo,
Bunkyo-ku, Tokyo 113-0033, Japan.}
\author{Masahito Hayashi}
\affiliation{Quantum Computation and Information Project, ERATO, JST, 5-28-3 Hongo,
Bunkyo-ku, Tokyo 113-0033, Japan..}
\pacs{03.67.-a,03.67.Hk}

\begin{abstract}
We propose a new protocol of \textit{universal} entanglement concentration,
which converts many copies of an \textit{unknown} pure state to an \textit{%
exact} maximally entangled state. The yield of the protocol, which is
outputted as a classical information, is probabilistic, and achieves the
entropy rate with high probability, just as non-universal entanglement
concentration protocols do.
\end{abstract}

\maketitle



\section{Introduction}

\label{s1} Conversion of a given partially entangled state to a maximally
entangled state by local operation and classical communication (LOCC) is an
important task in quantum information processing, both in application and
theory. If the given state is a pure state, such protocols are called
entanglement concentration, while mixed state versions are called
entanglement distillation.

In the paper, we study \textit{universal} entanglement concentration, or
entanglement concentration protocols which take unknown states as input, and
discuss the optimal yield in higher order asymptotic theory, or in
non-asymptotic theory, depending on the settings.

The reason why we studied universal entanglement concentration rather than
universal entanglement distillation is to study optimal yield in detail, in
comparison with non-universal protocols, which had been studied in detail ~%
\cite{Popescu,LP,Hardy,VidalJonathanNielsen,Morikoshi}. Note that study of
the optimal entanglement distillation is under development even in
non-universal settings. For example, the known formula for the optimal first
order rate still includes optimization over LOCC in one-copy space. This is
sharp contrast with the study of entanglement concentration, for which
non-asymptotic and higher order asymptotic formula in various form are
obtained.

As demonstrated by Bennett et al.~\cite{Ben}, if many copies of known pure
states are given, the optimal asymptotic yield equals the entropy of
entanglement of the input state \cite{Popescu}. To achieve the optimal, both
parties apply projections onto the typical subspaces of the reduced density
matrix of given partially entangled pairs (\textit{BBPS protocol},
hereafter). Obviously, the protocol is not applicable to the case where
information about Schmidt basis is unknown. Of course, one can estimate
necessary information by measuring some of the given copies. In such
protocol, however, the final state is not quite a maximally entangled state,
because errors in estimation of the Schmidt basis will cause distortions.

This paper proposes a protocol, denoted by $\left\{ C_{\ast }^{n}\right\} $,
of \textit{universal distortion-free entanglement concentration}, in which 
\textit{exact} (not approximate) entangled states are produced out of
identically prepared copies of an unknown pure state. Its yield is
probabilistic(, so users cannot predict the yield beforehand), but the
protocol outputs the amount of the yield as classical information, ( so that
users know what they have obtained, ) and the rate of the yield
asymptotically achieves the entropy of entanglement with probability close
to unity.

A key to construction of our protocol is symmetry; an ensemble of
identically prepared copies of a state is left unchanged by simultaneous
reordering of copies at each site. This symmetry gives rise to entanglement
which is accessible without any information about the Schmidt basis.

In some applications, small distortion in outputs might be enough, and
estimation-based protocols might suffice, because entropy rate is achieved
anyway. In higher order asymptotic terms and non-asymptotic evaluations,
however, we will prove that our protocol is better than any other protocols
which may allow small distortion.

In the proof, the following observation simplifies the problem to large
extent. Let us concentrate on the optimization of the worst-case quantity of
performance measures over all the unknown Schmidt basis, because the
uncertainty about Schmidt basis is the main difficulty of universal
entanglement concentration. We also assume performance measures are not
increasing by postprocessing which decreases the Schmidt rank of the product
maximally entangled state.

With such reasonable restrictions, an optimal protocol is always found out
in a class of protocols which are the same as $\left\{ C_{\ast }^{n}\right\} 
$ in output quantum states, but may differ in classical output. Therefore,
any trial of improvement of $\left\{ C_{\ast }^{n}\right\} $ cannot change
real yield. What can be 'improved' is the information about how much yield
was produced. For $\left\{ C_{\ast }^{n}\right\} $ outputs the information
about yield correctly, there should be no room for 'improvement' in this
part, too. This observation assures us that $\left\{ C_{\ast }^{n}\right\} $
is optimal if the criterion is fair. In addition, the optimization is now
straightforward, for we have to optimize only classical part of the protocol.

Based on this observation, we prove the optimality in terms of a natural
class of measures: monotone increasing measures which are bounded over the
range and continuously differentiable except at finitely many points. In
terms of such measures, distortion-free condition trivially implies the
non-asymptotic optimality of our protocol, while the constraint on the
distortion implies that our protocol is optimal up to the higher orders.
Also, (a kind of ) non-asymptotic optimality is proved for some performance
measures which varies with both yield and distortion. These results assure
us that our protocol $\{C_{\ast }^{n}\}$ is the best universal entanglement
concentration protocol.

Here, we stress that most of these results generalize to the case where
Schmidt coefficients of an input are known and its Schmidt basis is unknown.

In the end, we prove that the classical output of our protocol gives an
asymptotically optimal estimate of entropy of entanglement. Surprisingly,
this estimate is not less accurate than any other estimate based on
(potentially global) measurement whose construction depends on the Schmidt
basis of the unknown state.

Considering its optimal performance in very strong senses, it is surprising
that our protocol does not use any classical communication at all.

The paper is organized as follows. After introducing symbols and terms in
Section~\ref{sec:definition}, we describe implications of the permutation
symmetry, and constructed the protocol $\{C_{\ast }^{n}\}$ (Subsection~\ref%
{subsec:symmetry}). Its asymptotic performance is analyzed using known
results of group representation theoretic type theory in Subsection~\ref%
{sec:performance}, followed by comparison with a estimation based protocol~%
\ref{subsec:comparison-est-based}.

Optimality of the protocol is discussed in Section~\ref{sec:optimality}.
Subsection~\ref{subsec:whatisproved} gives definition of measures of
performance and short description of proved assertions. The key lemma, which
restricts the class of protocols of interest to large extent, is proved in
Subsection~\ref{subsec:keylemmas}. Subsections~\ref{subsec:dist-free}-\ref%
{subsec:totalfidelity} treats proof of optimality in each setting.

The estimation theoretic application of the protocol is discussed in Section~%
\ref{sec:estimate}. Interestingly, a part of arguments in this section gives
another proof of optimality of $\{C_{\ast }^{n}\}$ in terms of error
exponent.

In the appendices, we demonstrated several technical lemmas and formulas.
Among them, an asymptotic formula of average yield of non-universal
entanglement concentration protocols is, so far as we know, had not shown,
and might be useful for other applications.

\section{Definitions}

\label{sec:definition} Given an entangled pure state $|\phi \rangle \in 
\mathcal{H}_{A}\otimes \mathcal{H}_{B}$ ($\dim \mathcal{H}_{A}=\dim \mathcal{%
H}_{B}=d$), we denote its \textit{Schmidt coefficients} by $\mathbf{p}_{\phi
}=(p_{1,\phi },\ldots ,p_{d,\phi })$ ($p_{1,\phi }\geq p_{2,\phi }\geq
\ldots \geq p_{d,\phi }\geq 0)$ and its \textit{Schmidt basis} by $%
\{|e_{i,x}^{\phi }\}$, respectively. \textit{Entropy of entanglement} of $%
|\phi \rangle $ equals the Shannon entropy $\mathrm{H}$ of the probability
distribution $\mathbf{p}_{\phi }$, where Shannon entropy $\mathrm{H}$ is
defined by $\mathrm{H}(\mathbf{p}):=$ $\sum_{i}-p_{i}\log p_{i}$.
(Throughout the paper, the base of log is 2.) In the paper, our main concern
is concentration of maximal entanglement from $|\phi \rangle ^{\otimes n}$
by LOCC. We denote a maximally entangled state with the Schmidt rank $L$ by%
\begin{equation*}
\Vert L\rangle :=\frac{1}{\sqrt{L}}\sum_{i=1}^{L}|f_{i,A}^{n}\rangle
|f_{i,B}^{n}\rangle ,
\end{equation*}%
where $\{|f_{i,x}^{n}\rangle \}$ is an orthonormal basis in $\mathcal{H}%
_{x}^{\otimes n}(x=A,B)$. Note that $\{|f_{i,x}^{n}\rangle \}$ need not to
be explicitly defined, for the difference between $\frac{1}{\sqrt{L}}%
\sum_{i=1}^{L}|f_{i,A}^{n}\rangle |f_{i,B}^{n}\rangle $ and $\frac{1}{\sqrt{L%
}}\sum_{i=1}^{L}|\widetilde{f_{i,A}^{n}}\rangle |\widetilde{f_{i,B}^{n}}%
\rangle $ is compensated by a local unitary. One can optimally produce $%
\Vert 2^{n\mathrm{H}(\mathbf{p}_{\phi })}\rangle $ from $|\phi \rangle
^{\otimes n}$ by LOCC with high probability and high fidelity, if $n$ is
very large \cite{Ben,Popescu}.

In this paper, an entanglement concentration $\{C^{n}\}$ is a sequence of
LOCC measurement, in which $C^{n}$ takes $n$ copies $|\phi \rangle ^{\otimes
n}$ of unknown state as its input. With probability $Q_{C^{n}}^{\phi }(x)$, $%
C^{n}$ outputs $\rho _{C^{n}}^{\phi }(x)$, which is meant to be an
approximation to $\Vert 2^{nx}\rangle $, together with $x$ as classical
information.

The worst-case distortion $\epsilon _{C^{n}}^{\phi }$ is the maximum of
square of the Bure's distance between the output $\rho _{C^{n}}^{\phi }(x)$
and the target $\Vert 2^{nx}\rangle $, 
\begin{equation*}
\epsilon _{C^{n}}^{\phi }:=1-\min_{x}\langle 2^{nx}\Vert \rho _{C^{n}}^{\phi
}(x)\Vert 2^{nx}\rangle ,
\end{equation*}%
while $\bar{\epsilon}_{C^{n}}^{\phi }$ denotes the average distortion, 
\begin{eqnarray*}
\overline{\epsilon }_{C^{n}}^{\phi } &:&=1-\sum\limits_{x}Q_{C^{n}}^{\phi
}(x)\langle 2^{nx}\Vert \rho _{C^{n}}^{\phi }(x)\Vert 2^{nx}\rangle \\
&=&1-\mathrm{E}_{Q_{C^{n}}^{\phi }}^{X}\langle 2^{nX}\Vert \rho
_{C^{n}}^{\phi }(X)\Vert 2^{nX}\rangle ,
\end{eqnarray*}%
where $\mathrm{E}_{Q_{C^{n}}^{\phi }}^{X}$ means the average with respect to 
$Q_{C^{n}}^{\phi }$,%
\begin{equation*}
\mathrm{E}_{Q_{C^{n}}^{\phi }}^{X}f(X):=\sum\limits_{x}Q_{C^{n}}^{\phi
}(x)f(x).
\end{equation*}

A protocol is said to be \textit{distortion -free}, if $\epsilon
_{C^{n}}^{\phi }=\overline{\epsilon }_{C^{n}}^{\phi }=0$ holds for all $\phi 
$.

\section{Construction of the protocol $\{C_{\ast }^{n}\}$}

\label{sec:construction}

\subsection{Symmetry and the protocol $\{C_{\ast }^{n}\}$}

\label{subsec:symmetry} In the construction of $\{C_{\ast }^{n}\}$, we
exploit two kinds of symmetries. First, our input, $|\phi \rangle ^{\otimes
n}$, is invariant by the reordering of copies, or the action of the
permutation $\sigma $ in the set $\{1,\ldots n\}$ such that 
\begin{equation*}
\bigotimes_{i=1}^{n}|h_{i,A}\rangle |h_{i,B}\rangle \mapsto
\bigotimes_{i=1}^{n}|h_{\sigma ^{-1}(i),A}\rangle |h_{\sigma
^{-1}(i),B}\rangle ,
\end{equation*}%
\label{symmetric}where $|h_{i,x}\rangle \in \mathcal{H}_{x}\;(x=A,B)$.
(Hereafter, the totality of permutations in the set $\{1,...,n\}$ is denoted
by $S_{n}$.) Second, an action of local unitary transform $U^{\otimes n}{%
\otimes }V^{\otimes n}$ $(U,V\in \mathrm{SU}(d))$ corresponds to change of
the Schmidt basis.

Action of these groups induces a decomposition of the tensored space $%
\mathcal{H}_{x}^{\otimes n}(x=A,B)$ ~\cite{Weyl} into 
\begin{equation}
\mathcal{H}_{x}^{\otimes n}=\bigoplus_{\mathbf{n}}\mathcal{W}_{\mathbf{n}%
,x},\;\mathcal{W}_{\mathbf{n},x}:=\mathcal{U}_{\mathbf{n},x}\otimes \mathcal{%
V}_{\mathbf{n},x}\;(x=A,B),  \label{sym}
\end{equation}%
where $\mathcal{U}_{\mathbf{n},x}$ and $\mathcal{V}_{\mathbf{n},x}$ is an
irreducible space of the tensor representation of $\mathrm{SU}(d)$, and the
representation~(\ref{sym}) of the group of permutations respectively, and 
\begin{equation}
\mathbf{n}=(n_{1},\ldots ,n_{d}),\quad \sum_{i=1}^{d}n_{i}=n,\;n_{i}\geq
n_{i+1}\geq 0,  \label{yungindex}
\end{equation}%
is called \textit{Young index}, which $\mathcal{U}_{\mathbf{n},x}$ and $%
\mathcal{V}_{\mathbf{n},x}$ uniquely correspond to. In case of spin-$\frac{1%
}{2}$-system, $\mathcal{W}_{\mathbf{n},x}$ is an eigenspace of the total
spin operator. Due to the invariance by the permutation (\ref{sym}), any $n$%
-tensored state $|\phi \rangle ^{\otimes n}$ is decomposed in the following
form.

\begin{lem}
\begin{equation*}
|\phi \rangle ^{\otimes n}=\sum_{\mathbf{n}}\sqrt{a_{\mathbf{n}}^{\phi }}%
|\phi _{\mathbf{n}}\rangle \otimes |\mathcal{V}_{\mathbf{n}}\rangle ,
\end{equation*}%
where, $|\phi _{\mathbf{n}}\rangle $ is a \ state vector in $\mathcal{U}_{%
\mathbf{n},A}\otimes \mathcal{U}_{\mathbf{n},B}$, $a_{\mathbf{n}}^{\phi }$
is a complex number, and $|\mathcal{V}_{\mathbf{n}}\rangle $ is a maximally
entangled state in $\mathcal{V}_{\mathbf{n},A}\otimes \mathcal{V}_{\mathbf{n}%
,B}$ with the Schmidt rank $\dim \mathcal{V}_{\mathbf{n},A}.$ While $|\phi _{%
\mathbf{n}}\rangle $ and $a_{\mathbf{n}}^{\phi }$ depends on the input $%
|\phi \rangle $, $|\mathcal{V}_{\mathbf{n}}\rangle $ does not depend on the
input.
\end{lem}

\begin{pf}
Write 
\begin{equation*}
|\phi \rangle ^{\otimes n}=\sum_{\mathbf{n},\mathbf{n}^{\prime
}}\sum_{i,j,k,l}b_{\mathbf{n},i,j,\mathbf{n}^{\prime },k,l}\left\vert e_{i}^{%
\mathcal{U}_{\mathbf{n},A}}\right\rangle \left\vert e_{j}^{\mathcal{V}_{%
\mathbf{n},A}}\right\rangle \left\vert e_{k}^{\mathcal{U}_{\mathbf{n}%
^{\prime },B}}\right\rangle \left\vert e_{l}^{\mathcal{V}_{\mathbf{n}%
^{\prime },B}}\right\rangle
\end{equation*}%
where $\left\{ \left\vert e_{i}^{\mathcal{U}_{\mathbf{n},A}}\right\rangle
\right\} $, $\left\{ \left\vert e_{j}^{\mathcal{V}_{\mathbf{n}%
,A}}\right\rangle \right\} $, $\left\{ \left\vert e_{k}^{\mathcal{U}_{%
\mathbf{n}^{\prime },B}}\right\rangle \right\} $, $\left\{ \left\vert e_{l}^{%
\mathcal{V}_{\mathbf{n}^{\prime },B}}\right\rangle \right\} $ is a complete
orthonormal basis in $\mathcal{U}_{\mathbf{n},A}$, $\mathcal{V}_{\mathbf{n}%
,A}$, $\mathcal{U}_{\mathbf{n},B}$, $\mathcal{V}_{\mathbf{n},B}$,
respectively. Establish a correspondence between a vector $|\phi \rangle
^{\otimes n}$ in bipartite system and an operator 
\begin{equation*}
\Phi ^{n}:=\sum_{\mathbf{n},\mathbf{n}^{\prime }}\sum_{i,j,k,l}b_{\mathbf{n}%
,i,j,\mathbf{n}^{\prime },k,l}\left\vert e_{i}^{\mathcal{U}_{\mathbf{n}%
}}\right\rangle \left\langle e_{k}^{\mathcal{U}_{\mathbf{n}^{\prime
}}}\right\vert \otimes \left\vert e_{j}^{\mathcal{V}_{\mathbf{n}%
}}\right\rangle \left\langle e_{l}^{\mathcal{V}_{\mathbf{n}^{\prime
}}}\right\vert ,
\end{equation*}%
using 'partial transpose' or the linear map which maps $\left\vert e_{i}^{%
\mathcal{U}_{\mathbf{n},A}}\right\rangle \left\vert e_{j}^{\mathcal{V}_{%
\mathbf{n},A}}\right\rangle \left\vert e_{k}^{\mathcal{U}_{\mathbf{n}%
^{\prime },B}}\right\rangle \left\vert e_{l}^{\mathcal{V}_{\mathbf{n}%
^{\prime },B}}\right\rangle $ to $\left\vert e_{i}^{\mathcal{U}_{\mathbf{n}%
}}\right\rangle \left\langle e_{k}^{\mathcal{U}_{\mathbf{n}^{\prime
}}}\right\vert \otimes \left\vert e_{j}^{\mathcal{V}_{\mathbf{n}%
}}\right\rangle \left\langle e_{l}^{\mathcal{V}_{\mathbf{n}^{\prime
}}}\right\vert $. For this map is one to one, we study $\Phi ^{n}$ in stead
of $|\phi \rangle ^{\otimes n}$.

Observe that $\Phi ^{n}$ is invariant by action of any permutation $\sigma $%
, 
\begin{equation*}
\sigma \Phi ^{n}\sigma ^{\dagger }=\Phi ^{n},
\end{equation*}%
where the action of $\sigma $ is defined by (\ref{symmetric}). Due to Lemma %
\ref{lem:decohere}, $b_{\mathbf{n},i,j,\mathbf{n}^{\prime },k,l}=0$ unless $%
\mathbf{n}=\mathbf{n}^{\prime }$, and%
\begin{equation*}
\Phi ^{n}=\bigoplus_{\mathbf{n}}\sum_{i,j,k,l}b_{\mathbf{n},i,j,\mathbf{n}%
,k,l}\left\vert e_{i}^{\mathcal{U}_{\mathbf{n}}}\right\rangle \left\langle
e_{k}^{\mathcal{U}_{\mathbf{n}}}\right\vert \otimes \left\vert e_{j}^{%
\mathcal{V}_{\mathbf{n}}}\right\rangle \left\langle e_{l}^{\mathcal{V}_{%
\mathbf{n}}}\right\vert .
\end{equation*}%
Then, we apply Lemma \ref{lem:shur2} $\ $to 
\begin{equation*}
\sum_{i,j,k,l}b_{\mathbf{n},i,j,\mathbf{n},k,l}\left\vert e_{i}^{\mathcal{U}%
_{\mathbf{n}}}\right\rangle \left\langle e_{k}^{\mathcal{U}_{\mathbf{n}%
}}\right\vert \otimes \left\vert e_{j}^{\mathcal{V}_{\mathbf{n}%
}}\right\rangle \left\langle e_{l}^{\mathcal{V}_{\mathbf{n}}}\right\vert ,
\end{equation*}%
proving that $\Phi ^{n}$ is of the form 
\begin{eqnarray*}
\Phi ^{n} &=&\bigoplus_{\mathbf{n}}\sum_{i,k}b_{\mathbf{n},i,j}^{\prime
}\left\vert e_{i}^{\mathcal{U}_{\mathbf{n}}}\right\rangle \left\langle
e_{k}^{\mathcal{U}_{\mathbf{n}}}\right\vert \otimes \mathrm{Id}_{\mathcal{V}%
_{\mathbf{n}}} \\
&=&\bigoplus_{\mathbf{n}}\sqrt{a_{\mathbf{n}}^{\phi }}\Phi _{\mathbf{n}%
}\otimes \sqrt{\frac{1}{\dim \mathcal{V}_{\mathbf{n}}}}\sum_{j=1}^{\dim 
\mathcal{V}_{\mathbf{n}}}\left\vert e_{j}^{\mathcal{V}_{\mathbf{n}%
}}\right\rangle \left\langle e_{j}^{\mathcal{V}_{\mathbf{n}}}\right\vert ,
\end{eqnarray*}%
where $\Phi _{\mathbf{n}}$ is a linear map in $\mathcal{U}_{\mathbf{n}}$. To
obtain the lemma, we simply take "partial transpose" of this again: apply
the linear map which maps $\left\vert e_{i}^{\mathcal{U}_{\mathbf{n}%
}}\right\rangle \left\langle e_{k}^{\mathcal{U}_{\mathbf{n}^{\prime
}}}\right\vert \otimes \left\vert e_{j}^{\mathcal{V}_{\mathbf{n}%
}}\right\rangle \left\langle e_{l}^{\mathcal{V}_{\mathbf{n}^{\prime
}}}\right\vert $ to $\left\vert e_{i}^{\mathcal{U}_{\mathbf{n}%
,A}}\right\rangle \left\vert e_{j}^{\mathcal{V}_{\mathbf{n},A}}\right\rangle
\left\vert e_{k}^{\mathcal{U}_{\mathbf{n}^{\prime },B}}\right\rangle
\left\vert e_{l}^{\mathcal{V}_{\mathbf{n}^{\prime },B}}\right\rangle $. For
this map is one to one, $\Phi ^{n}$ is mapped to $|\phi \rangle ^{\otimes n}$%
. By this map, $\sqrt{\frac{1}{\dim \mathcal{V}_{\mathbf{n}}}}%
\sum_{j=1}^{\dim \mathcal{V}_{\mathbf{n}}}\left\vert e_{j}^{\mathcal{V}_{%
\mathbf{n}}}\right\rangle \left\langle e_{j}^{\mathcal{V}_{\mathbf{n}%
}}\right\vert $ is mapped to $|\mathcal{V}_{\mathbf{n}}\rangle $, and $\Phi
_{\mathbf{n}}$ is mapped to $|\phi _{\mathbf{n}}\rangle \in \mathcal{U}_{%
\mathbf{n},A}\otimes \mathcal{U}_{\mathbf{n},B}$.
\end{pf}

This lemma implies that there are maximally entangled states, $|\mathcal{V}_{%
\mathbf{n}}\rangle $, which are accessible without using knowledge on the
input state. The average amount of the accessible entanglement is decided by
the coefficients $a_{\mathbf{n}}^{\phi }$, which vary with the Schmidt
coefficients of the input $|\phi \rangle $.

Now we are at the position to present our universal distortion-free
entanglement concentration protocol $\left\{ C_{\ast }^{n}\right\} $
(Hereafter, the projection onto a Hilbert space $\mathcal{X}$ is also
denoted by $\mathcal{X}$): First, each party apply the projection
measurements $\{\mathcal{W}_{\mathbf{n}_{A},A}\}_{\mathbf{n}_{A}}$ and $\{%
\mathcal{W}_{\mathbf{n}_{B},B}\}_{\mathbf{n}_{B}}$ at each site
independently. This yields the same measurement result $\mathbf{n}_{A}=%
\mathbf{n}_{B}=\mathbf{n}$ at both site, and the state is changed to $|\phi
_{\mathbf{n}}\rangle \otimes |\mathcal{V}_{\mathbf{n}}\rangle $. Taking
partial trace over $\mathcal{U}_{\mathbf{n},A}$ and $\mathcal{U}_{\mathbf{n}%
,B}$ at each site, we obtain $|\mathcal{V}_{\mathbf{n}}\rangle $.

If $\mathcal{H}_{A}$ and $\mathcal{H}_{B}$ are qubit systems, $\{\mathcal{W}%
_{\mathbf{n}_{A},A}\}_{\mathbf{n}_{A}}$ is nothing but the measurement of
the total angular momentum.

For the sake of the formalism, $|\mathcal{V}_{\mathbf{n}}\rangle $ is mapped
to $\Vert \dim \mathcal{V}_{\mathbf{n}}\rangle $. With this modification, $%
\rho _{C_{\ast }^{n}}^{\phi }(x)=\Vert 2^{nx}\rangle \langle 2^{nx}\Vert $
and $Q_{C_{\ast }^{n}}^{\phi }(x)=a_{\mathbf{n}}^{\phi }$, if $2^{nx}=\dim 
\mathcal{V}_{\mathbf{n}}$ (if such $\mathbf{n}$ does not exist, $Q_{C_{\ast
}^{n}}^{\phi }(x)=0$).

Due to the identity $Q_{C_{\ast }^{n}}^{\phi }\left( \frac{\log \dim 
\mathcal{V}_{\mathbf{n}}}{n}\right) =a_{\mathbf{n}}^{\phi }=\mathrm{Tr}%
\left\{ \mathcal{W}_{\mathbf{n},A}\left( \mathrm{Tr}_{B}|\phi \rangle
\langle \phi |\right) ^{\otimes n}\right\} $ and the formulas in the
appendix of \cite{Ha}, we can evaluate the asymptotic behavior of $%
Q_{C_{\ast }^{n}}^{\phi }(x)$ as follows:

\begin{equation}
\begin{split}
\left\vert \frac{\log \dim \mathcal{V}_{\mathbf{n}}}{n}-\mathrm{H}\left( 
\frac{\mathbf{n}}{n}\right) \right\vert & \leq \frac{d^{2}+2d}{2n}\log (n+d),
\\
\lim_{n\rightarrow \infty }\frac{-1}{n}\log Q_{C_{\ast }^{n}}^{\phi }\left( 
\frac{\log \dim \mathcal{V}_{\mathbf{n}}}{n}\right) & =\mathrm{D}(\frac{%
\mathbf{n}}{n}\Vert \mathbf{p}_{\phi }), \\
\lim_{n\rightarrow \infty }\frac{-1}{n}\log \sum_{\frac{\mathbf{n}}{n}\in 
\mathcal{R}}Q_{C_{\ast }^{n}}^{\phi }\left( \frac{\log \dim \mathcal{V}_{%
\mathbf{n}}}{n}\right) & =\max_{\mathbf{q}\in \mathcal{R}}\mathrm{D}(\mathbf{%
q}\Vert \mathbf{p}_{\phi }),
\end{split}
\label{grep-type}
\end{equation}%
where $\mathcal{R}$ is an arbitrary closed subset of $\{\mathbf{q}|q_{1}\geq
q_{2}\geq \ldots \geq q_{d}\geq 0,\sum_{i=1}^{d}q_{i}=1\}$. These means the
probability for $\frac{1}{n}\log \dim \mathcal{V}_{\mathbf{n}}\sim \mathrm{H}%
\left( \boldsymbol{p}\right) $ is exponentially close to unity, as is
demonstrated in Subsection \ref{sec:performance}.

\subsection{Asymptotic Performance of $\{C_{\ast }^{n}\}$}

\label{sec:performance} In this subsection, we analyze the asymptotic
performance of $\{C_{\ast }^{n}\}$ in terms of success (failure)
probability, total fidelity, and average of the log of Schmidt rank of the
output maximally entangled states. (The proof of the optimality of $%
\{C_{\ast }^{n}\}$ is made for more general class of measures). For the main
difficulty of universal concentration is attributed to uncertainty about
Schmidt basis, we consider the value in the worst-case Schmidt basis.

The worst-case value for the failure probability, or the probability that
the yield is not more than $y$ equals 
\begin{equation}
\max_{U,V}\sum\limits_{x:x\leq y}Q_{C^{n}}^{U\otimes V\phi }(x).
\label{error-prob}
\end{equation}%
where $U$ and $V$ run all over unitary matrices. For the yield of our
protocol $\{C_{\ast }^{n}\}$ is invariant by local unitary operations, the
maximum over $U$ and $V$ can be removed. Due to the first and the third
formula in (\ref{grep-type}), we have,

\begin{equation}
\lim_{n\rightarrow \infty }\frac{-1}{n}\log \sum\limits_{x:x\leq
R}Q_{C_{\ast }^{n}}^{\phi }(x)=\mathrm{D}(R\Vert \mathbf{p}_{\phi }),
\label{fexp1-1}
\end{equation}%
and 
\begin{equation}
\lim_{n\rightarrow \infty }\frac{-1}{n}\log \sum\limits_{x:x\geq
R}Q_{C_{\ast }^{n}}^{\phi }(x)=\mathrm{D}(R\Vert \mathbf{p}_{\phi })
\label{fexp-2}
\end{equation}%
where

\begin{equation*}
\mathrm{D}(R\Vert \mathbf{p})=\left\{ 
\begin{array}{cc}
\min_{\mathbf{q}:\,H(\mathbf{q})\geq R}\mathrm{D}(\mathbf{q}\Vert \mathbf{p})
& (\mathrm{H}(\mathbf{p})\leq R), \\ 
\min_{\mathbf{q}:\,H(\mathbf{q})\leq R}\mathrm{D}(\mathbf{q}\Vert \mathbf{p})
& (\mathrm{H}(\mathbf{p})>R).%
\end{array}%
\right.
\end{equation*}%
\ 

Eq.~(\ref{fexp1-1}) implies that our protocol achieves entropy rate: if $R$
is strictly smaller than $\mathrm{H}(\mathbf{p}_{\phi })$, the RHS of \ Eq.~(%
\ref{fexp1-1}) is positive, which means that the failure probability is
exponentially small. On the other hand, Eq.~(\ref{fexp-2}) means that the
probability to have the yield more than the optimal rate (\textit{strong
converse probability}) tends to vanish, and its convergence is exponentially
fast.

Next, we evaluate the exponent of the \textit{total fidelity} $%
F_{C^{n}}^{\phi }\left( R\right) $, or the average fidelity to the maximally
entangled state whose Schmidt rank is not smaller than $2^{nR}$: 
\begin{equation}
F_{C^{n}}^{\phi }\left( R\right) :=\mathrm{E}_{Q_{C^{n}}^{\phi
}}^{X}\max_{y:y\geq R}\langle 2^{ny}\Vert \rho _{C^{n}}^{\phi }(X)\Vert
2^{ny}\rangle .  \label{def:totalfidelity}
\end{equation}%
(The optimization is considered in the worst-case Schmidt basis.) This
function describes trade-off between yield and distortion. Obviously, $%
F_{C^{n}}^{\phi }\left( R\right) $ is non-increasing in $R$, and takes
larger value if the protocol is better. We evaluate this quantity for $%
\{C_{\ast }^{n}\}$ as follows. 
\begin{eqnarray*}
&&1-F_{C_{\ast }^{n}}^{\phi }\left( R\right) =1-\sum_{x}\min \left\{
1,2^{-n(R-x)}\right\} Q_{C_{\ast }^{n}}^{\phi }(x) \\
&=&\sum_{x:x<R}\left( 1-2^{-n(R-x)}\right) Q_{C_{\ast }^{n}}^{\phi }(x).
\end{eqnarray*}%
The RHS is upper-bounded by $\sum_{x:x<R}Q_{C_{\ast }^{n}}^{\phi }(x)$ and
lower-bounded by $\left( 1-2^{-n(R-x)}\right) Q_{C_{\ast }^{n}}^{\phi }(x)$
where $x$ can be any value strictly smaller than $R$. Hence, if $R<H(\mathbf{%
p}_{\phi })$ letting $x=R-\frac{c}{n}$ such that $Q_{C_{\ast }^{n}}^{\phi
}(x)\neq 0$, using the second equation of (\ref{grep-type}), we have 
\begin{equation*}
\lim_{n\rightarrow \infty }\frac{-1}{n}\log \left( 1-F_{C_{\ast }^{n}}^{\phi
}\left( R\right) \right) =\mathrm{D}(R\Vert \mathbf{p}_{\phi }).
\end{equation*}

The exponent of failure probability, strong converse probability, and total
fidelity for the optimal non-universal protocol are found out in \cite%
{Morikoshi}, and we can observe the non-zero gap between the exponents of $%
\left\{ C_{\ast }^{n}\right\} $ and the optimal non-universal protocol. By
contrast, these quantities for BBPS protocol coincides with the ones for $%
\left\{ C_{\ast }^{n}\right\} $. (Proof is straightforwardly done using the
classical type theory).

This fact may imply that the protocol $\left\{ C_{\ast }^{n}\right\} $ is so
well-designed that its performance is comparable with the one which uses
some information about the input state. However, it might be the case that
these quantities are not sensitive to difference in performance. Hence, we
also discuss another quantity, the average yield (, evaluated at the
worst-case Schmidt basis), 
\begin{equation}
\min_{U,V}\sum_{x}xQ_{C^{n}}^{U\otimes V\phi }(x)=\min_{U,V}\mathrm{E}%
_{Q_{C^{n}}^{U\otimes V\phi }}^{X}X.  \label{average-yield}
\end{equation}%
The average yield of BBPS protocol is of the form%
\begin{equation*}
\mathrm{H}(\mathbf{p}_{\phi })+A\frac{\log n}{n}+\frac{B}{n}+o\left( \frac{1%
}{n}\right) ,
\end{equation*}%
where the coefficients $A$, $B$ and their derivation are described in
Appendix \ref{sec:cal-ave-ben}. The average yield of the protocol $\left\{
C_{\ast }^{n}\right\} $ is less than that of BBPS protocol by $\frac{C}{n}$,
where $C$ is calculated in Appendix~\ref{sec:diff-ave-yield}. Hence, this
measure is sensitive to the difference in performance which do not reveal in
the exponent of failure probability etc.

\subsection{Comparison with estimation based protocols}

\label{subsec:comparison-est-based}Most straightforwardly, universal
entanglement concentration is constructed based on the state estimation;
First, $c_{n}$ copies of $\left\vert \phi \right\rangle $ are used to
estimate the Schmidt basis, and second, apply BBPS protocol to the $n-c_{n}$
copies of $\left\vert \phi \right\rangle $. The average yield of such
protocol cannot be better than

\begin{equation*}
\frac{n-c_{n}}{n}\mathrm{H}\left( \mathbf{p}\right) +A\frac{\log (n-c_{n})}{%
n-c_{n}}+O\left( \frac{1}{n}\right) ,
\end{equation*}%
where $c_{n}$ slowly grows as $n$ increases. Therefore, this
estimation-based protocol cannot be better than $\left\{ C_{\ast
}^{n}\right\} $, because the average yield of $\left\{ C_{\ast }^{n}\right\} 
$ and BBPS protocol are the same except for $O\left( \frac{1}{n}\right) $%
-terms.

One might improve the estimation-based protocol by replacing BBPS protocol
with the non-asymptotically optimal entanglement concentration protocol.
However, this improvement is not likely to be effective, because in qubit
case, the yield of these protocols are the same up to the order of $O\left( 
\frac{\log n}{n}\right) $ (Appendix \ref{sec:cal-ave-opt}, $O\left( \frac{1}{%
n}\right) $-term is also given).

Another alternative is to use precise measurements which cause only
negligible distortion, so that we can use all the given copies of an unknown
state for entanglement concentration. This protocol can be very good, and
there might be many other good protocols. As is proven in the next section,
however, none of these protocols is no better than $\left\{ C_{\ast
}^{n}\right\} $, i.e., $\left\{ C_{\ast }^{n}\right\} $ is optimal for all
protocols whose outputs are slightly distorted.

\section{Optimality of $\{C_{\ast }^{n}\}$}

\label{sec:optimality}

\subsection{Measures, settings, and summary of results}

\label{subsec:whatisproved}A performance of an entanglement concentration
has two parts. One is amount of yield, and the other is distortion of the
output. The measures of the latter are, as is explained in Section\thinspace %
\ref{sec:definition}, $\epsilon _{C^{n}}^{\phi }$ and $\overline{\epsilon }%
_{C^{n}}^{\phi }$. Hereafter, maximum of these quantities over all Schmidt
basises ( $\max_{U,V}\epsilon _{C^{n}}^{U\otimes V\phi }$ and $\max_{U,V}%
\overline{\epsilon }_{C^{n}}^{U\otimes V\phi }$) are discussed.

The measures of the yield (\ref{error-prob}), (\ref{average-yield})
discussed in the previous section are essentially of the form 
\begin{equation}
\min_{U,V}\sum_{x}f\left( x\right) Q_{C^{n}}^{U\otimes V\phi }(x)=\min_{U,V}%
\mathrm{E}_{Q_{C^{n}}^{U\otimes V\phi }}^{X}f\left( X\right) .
\label{g-yield}
\end{equation}%
So far, we had considered minimization for error probability, and
maximization for average yield. Hereafter, we use success probability%
\begin{equation*}
\min_{U,V}\mathrm{E}_{Q_{C^{n}}^{U\otimes V\phi }}^{X}\Theta \left(
X-R\right) ,
\end{equation*}%
with $\Theta (x)$ denoting the step function, instead of error probability (%
\ref{error-prob}). From here to the end, optimization of an yield measure (%
\ref{g-yield}) means maximization of (\ref{g-yield}).

Namely, minimization of (\ref{error-prob}) corresponds to maximization of (%
\ref{g-yield}) with $f(x)=\Theta (x-R)$. Also, maximization of (\ref%
{average-yield}) is equivalent to maximization of (\ref{g-yield}) with $f(x)=%
\frac{x}{\log d}$.

These examples are monotone and bounded, or 
\begin{eqnarray}
&&f(x)\geq f(x^{\prime })\geq f(0)=0,\quad (x\geq x^{\prime }\geq 0),
\label{cond-f} \\
&&f(\log d)=1,  \label{cond-f-2}
\end{eqnarray}%
and%
\begin{equation}
\mbox{continuously differentiable but finitely many points.}
\label{cond-f-3}
\end{equation}%
The condition (\ref{cond-f}) and (\ref{cond-f-2}) are assumed throughout the
paper unless otherwise mentioned.

In the following subsections, measures of the form (\ref{g-yield}) are
optimized with the restriction on the worst-case distortion $%
\max_{U,V}\epsilon _{C^{n}}^{U\otimes V\phi }$ or the average distortion $%
\max_{U,V}\overline{\epsilon }_{C^{n}}^{U\otimes V\phi }$.

Also, we consider the optimization (maximization) of the measures which vary
with both yield and distortion. Namely, the weighted sum of these yield
measure and the average distortion $\overline{\epsilon }_{C^{n}}^{\phi }$,
i.e., 
\begin{equation}
\min_{U,V}\mathrm{E}_{Q_{C^{n}}^{U\otimes V\phi }}^{X}f(X)-\lambda \max_{U,V}%
\overline{\epsilon }_{C^{n}}^{U\otimes V\phi },  \label{g-y+d}
\end{equation}%
and the total fidelity $\max_{U,V}F_{C^{n}}^{U\otimes V\phi }\left( R\right) 
$ are considered.

We prove that the protocol is optimal, in the following senses.

\begin{enumerate}
\item The entropy rate is achieved (Subsection \ref{sec:performance}, Eq. (%
\ref{fexp1-1})).

\item Non-asymptotic behavior is best among all distortion-free protocols
(Subsection \ref{subsec:dist-free}).

\item Higher order asymptotic behavior is best for all protocols which
allows small distortions (Subsections \ref{subsec:worst-distortion}, \ref%
{subsec:average-distortion}).

\item In terms of weighed sum measures (\ref{g-y+d}) and the total fidelity,
the non-asymptotic optimality holds (Subsections \ref{subsec:weighted}-\ref%
{subsec:totalfidelity}).
\end{enumerate}

The key to the proof of these assertions is Lemma~\ref{lem:k1} which will be
proved in the next section. Due to this lemma, we can focus on the protocols
which is a modification of $\{C_{\ast }^{n}\}$ in its classical outputs
only. This fact not only simplifies the argument but also assures us that $%
\{C_{\ast }^{n}\}$ is a very natural protocol.

Here, we note that many of our results in this section generalize to the
case where Schmidt basis is unknown and Schmidt coefficients are known. Such
generalization is possible if the optimization problem can be recasted only
in terms of a family of quantum states $\left\{ U\otimes V\phi \right\}
_{UV} $, where $\left\vert \phi \right\rangle $ is a given input state and $%
U $ and $V$ run all over $\mathrm{SU}(d)$. This is trivially the case when
we optimize a function of distortion and yield. This is also the case if
conditions on distortion are needed to be imposed only on the given input
state, and not on all the possible input states.

\subsection{The Key Lemma}

\label{subsec:keylemmas}\vspace*{-2pt}In this subsection, we prove Lemma \ref%
{lem:k1}, which is the key to the arguments in the rest of the paper. To
make analysis easier, before the protocol starts, each party applies $%
U^{\otimes n}$, $V^{\otimes n}$ at each site, where $U$ and $V$ are chosen
randomly according to Haar measure in $\mathrm{SU}(d)$, and erase the memory
of $U,V$. This operation is denoted by $O1$, hereafter.

From here to the end of the paper, $C_{\ast }^{n}$ means the composition of $%
O1$ followed by $C_{\ast }^{n}$. The optimality of the newly defined $%
\left\{ C_{\ast }^{n}\right\} $ trivially implies the optimality of $\left\{
C_{\ast }^{n}\right\} $ defined previously, because $O1$ simply randomizes
output and cannot improve the performance;%
\begin{equation*}
O1:\rho \rightarrow \mathrm{E}_{U,V}(U\otimes V)^{\otimes n}\rho (U^{\dagger
}\otimes V^{\dagger })^{\otimes n},
\end{equation*}%
where $\mathrm{E}_{U,V}$ denotes expectation by Haar measure in $\mathrm{SU}%
(d)$. (More explicitly,%
\begin{equation*}
\mathrm{E}_{U,V}f(U,V)=\int f(U,V)\mu (\mathrm{d}U)\mu (\mathrm{d}V),
\end{equation*}%
where $\mu $ is the Haar measure with the convention $\int \mu (\mathrm{d}%
U)=1$.)

Lemma~\ref{l5} implies that $\mathcal{U}_{\mathbf{n},A}\otimes \mathcal{U}_{%
\mathbf{n},B}$ is an irreducible space of the tensored representation $%
U^{\otimes n}\otimes V^{\otimes n}$ of $\mathrm{SU}(d)\times \mathrm{SU}(d)$%
. Hence, by virtue of Lemmas~\ref{lem:decohere}-\ref{lem:shur}, the average
state writes%
\begin{equation}
\mathrm{E}_{U,V}(U\otimes V|\phi \rangle \langle \phi |U^{\ast }\otimes
V^{\ast })^{\otimes n}=\bigoplus_{\mathbf{n}}a_{\mathbf{n}}^{\phi }\sigma _{%
\mathbf{n}}^{\phi },  \label{5-5-1}
\end{equation}%
and%
\begin{equation*}
\sigma _{\mathbf{n}}^{\phi }:=\frac{\mathcal{U}_{\mathbf{n},A}\otimes 
\mathcal{U}_{\mathbf{n},B}\otimes |\mathcal{V}_{\mathbf{n}}\rangle \langle 
\mathcal{V}_{\mathbf{n}}|}{\dim \left\{ \mathcal{U}_{\mathbf{n},A}\otimes 
\mathcal{U}_{\mathbf{n},B}\right\} }.
\end{equation*}

We denote by $O2$ the projection measurement $\{\mathcal{W}_{\mathbf{n}%
_{A},A}\otimes \mathcal{W}_{\mathbf{n}_{B},B}\}_{\mathbf{n}_{A},\mathbf{n}%
_{B}}$,\textit{\ }which maps the state $\rho $ to the pair 
\begin{equation*}
\left( \mathbf{n}_{A},\,\mathbf{n}_{B},\,\frac{\mathcal{W}_{\mathbf{n}%
_{A},A}\otimes \mathcal{W}_{\mathbf{n}_{B},B}\rho \mathcal{W}_{\mathbf{n}%
_{A},A}\otimes \mathcal{W}_{\mathbf{n}_{B},B}}{\mathrm{tr}\mathcal{W}_{%
\mathbf{n}_{A},A}\otimes \mathcal{W}_{\mathbf{n}_{B},B}\rho }\right) 
\end{equation*}%
with probability 
\begin{equation*}
\mathrm{tr}\mathcal{W}_{\mathbf{n}_{A},A}\otimes \mathcal{W}_{\mathbf{n}%
_{B},B}\rho .
\end{equation*}%
Here note that, due to the form of $\sigma _{\mathbf{n}}^{\phi }$, $\mathbf{n%
}_{A}=\mathbf{n}_{B}:=\mathbf{n}$, so long as the input is many copies of a
pure state. Given a pair $\left( \mathbf{n},\sigma _{\mathbf{n}}^{\phi
}\right) $ of classical information and a state supported on $\mathcal{W}_{%
\mathbf{n},A}\otimes \mathcal{W}_{\mathbf{n},B}$, the operation $O3$ outputs 
$\left( \mathbf{n},\mathrm{tr}_{\mathcal{U}_{\mathbf{n},A}\otimes \mathcal{U}%
_{\mathbf{n},B}}\sigma _{\mathbf{n}}^{\phi }\right) $.

Denoting the composition of an operation $A$ followed by an operation $B$ as 
$B\circ A$, $C_{\ast }^{n}$ writes $O3\circ O2\circ O1$, essentially. (The
mapping from $|\mathcal{V}_{\mathbf{n}}\rangle $ to $||\dim \mathcal{V}_{%
\mathbf{n}}\rangle $ is needed only for the sake of formality.) Here, in
defining $B\circ A$, if $A$'s output is a pair $(\mathbf{n},\rho _{\mathbf{n}%
})$ of classical information and quantum state, we always consider the
correspondence

\begin{equation}
(\mathbf{n},\rho _{\mathbf{n}})\leftrightarrow \left\vert \mathbf{n}%
\right\rangle \left\langle \mathbf{n}\right\vert \otimes U_{\mathbf{n}}\rho
_{\mathbf{n}}U_{\mathbf{n}}^{\dagger },  \label{correspondence}
\end{equation}%
where $\{\left\vert \mathbf{n}\right\rangle \}$ is an orthonormal basis, and 
$U_{x}$ is an local isometry to appropriately defined Hilbert space. Here,
'local' is in terms of A-B partition. In terms of this convention, the
definition of $O3$ rewrites%
\begin{equation}
O3:\left\vert \mathbf{n}\right\rangle \left\langle \mathbf{n}\right\vert
\otimes U_{\mathbf{n}}\sigma _{\mathbf{n}}^{\phi }U_{\mathbf{n}}^{\dagger
}\rightarrow \left\vert \mathbf{n}\right\rangle \left\langle \mathbf{n}%
\right\vert \otimes U_{\mathbf{n}}^{\prime }\mathrm{tr}_{\mathcal{U}_{%
\mathbf{n},A}\otimes \mathcal{U}_{\mathbf{n},B}}\sigma _{\mathbf{n}}^{\phi
}U_{\mathbf{n}}^{\prime \dagger },  \label{defUn}
\end{equation}%
using local isometry $U_{\mathbf{n}}$ and $U_{\mathbf{n}}^{\prime }$. (The
domain of $U_{\mathbf{n}}$ and $U_{\mathbf{n}}^{\prime }$is $\mathcal{W}_{%
\mathbf{n},A}\otimes \mathcal{W}_{\mathbf{n},B}$ and $\mathcal{V}_{\mathbf{n}%
,A}\otimes \mathcal{V}_{\mathbf{n},B}$, respectively.)

Recall that all the measures listed in the previous section are invariant by
local unitary operations to the input, i.e., the measure $f_{n}\left( \rho
,\{C^{n}\}\right) $ satisfies 
\begin{eqnarray}
f_{n}\left( \rho ,\{C^{n}\}\right)  &=&f_{n}\left( U\otimes V\rho U^{\dagger
}\otimes V^{\dagger },\{C^{n}\}\right) ,  \label{inv-measure} \\
\forall U,\forall V &\in &\mathrm{SU}(d),  \notag
\end{eqnarray}%
and is affine with respect to $\rho $,%
\begin{eqnarray}
&&f_{n}\left( p\rho +(1-p)\sigma ,\{C^{n}\}\right)   \label{affine-measure}
\\
&=&pf_{n}\left( \rho ,\{C^{n}\}\right) +(1-p)f_{n}\left( \sigma
,\{C^{n}\}\right) .  \notag
\end{eqnarray}%
Recall also that the worst-case/average distortion are affine. Hereafter,
the worst-case/average distortion are always evaluated at the worst-case
Schmidt basis, so that those measures satisfy (\ref{inv-measure}).

\begin{lem}
For any given protocol $\{C^{n}\}$, we can find a protocol such that; (i)
The protocol is of the form $\{B^{n}\circ C_{\ast }^{n}\}$, where $B^{n}$ is
an LOCC operation; (ii) A performance measure satisfying (\ref{inv-measure})
and (\ref{affine-measure})takes the same value as the protocol $\{C^{n}\}$,%
\begin{equation*}
f(\rho ,\left\{ B^{n}\circ C_{\ast }^{n}\right\} )=f(\rho ,\left\{
C^{n}\right\} ).
\end{equation*}%
\label{optimal}
\end{lem}

\begin{pf}
Due to (\ref{inv-measure}) and (\ref{affine-measure}), the operation $O1$
does not decrease the measure of the performance, because 
\begin{eqnarray*}
&&f_{n}\left( \mathrm{E}_{U,V}U\otimes V\rho U^{\dagger }\otimes V^{\dagger
},\{C^{n}\}\right) \\
&=&\mathrm{E}_{U,V}\,f_{n}\left( U\otimes V\rho U^{\dagger }\otimes
V^{\dagger },\{C^{n}\}\right) \\
&=&\mathrm{E}_{U,V}\,f_{n}\left( \rho ,\{C^{n}\}\right) =\,f_{n}\left( \rho
,\{C^{n}\}\right) .
\end{eqnarray*}%
Hence, $\{C^{n}\circ O1\}$ is the same as $\{C^{n}\}$ in the performance.

After the operation $O1$, the state is block diagonal in subspaces $\{%
\mathcal{W}_{\mathbf{n},A}\otimes \mathcal{W}_{\mathbf{n},B}\}$. Therefore,
if we use the correspondence (\ref{correspondence}), the state is not
unchanged by $O2$ (up to local isometry). More explicitly, let $U_{\mathbf{n}%
}$ be a local isometry in (\ref{defUn}), and $C^{\prime n}=C^{n}\circ U_{%
\mathbf{n}}^{\dagger }$. Then, we have 
\begin{equation*}
f(\rho ,\left\{ C^{\prime n}\circ O2\circ O1\right\} )=f(\rho ,\left\{
C^{n}\circ O1\right\} ).
\end{equation*}

Observe also that, after the operation $O1$, parts of the state which are
supported on $\mathcal{U}_{\mathbf{n},A}\otimes \mathcal{U}_{\mathbf{n},B}$
are tensor product states. Hence, there is a operation $B^{n}$ such that 
\begin{equation*}
f(\rho ,\left\{ B^{n}\circ O3\circ O2\circ O1\right\} )=f(\rho ,\left\{
C^{\prime n}\circ O2\circ O1\right\} ),
\end{equation*}%
because tensor product states can be reproduced locally whenever they are
needed. More explicitly, $B^{n}=C^{\prime n}\circ B^{\prime n}$, where $%
B^{\prime n}$ is 
\begin{equation*}
B^{\prime n}:\left\vert \mathbf{n}\right\rangle \left\langle \mathbf{n}%
\right\vert \otimes U_{\mathbf{n}}^{\prime }\rho U_{\mathbf{n}}^{\prime
\dagger }\rightarrow \left\vert \mathbf{n}\right\rangle \left\langle \mathbf{%
n}\right\vert \otimes U_{\mathbf{n}}\left( \mathcal{U}_{\mathbf{n},A}\otimes 
\mathcal{U}_{\mathbf{n},B}\otimes \rho \right) U_{\mathbf{n}}^{\dagger },
\end{equation*}%
where $U_{\mathbf{n}}$ and $U_{\mathbf{n}}^{\prime }$ are local isometry in (%
\ref{defUn}).

After all,$f(\rho ,\left\{ B^{n}\circ C_{\ast }^{n}\right\} )=f(\rho
,\left\{ C^{n}\right\} )$ and the lemma is proved.
\end{pf}

In the postprocessing $B^{n}$, a classical output $x$ will be changed to $%
x+\Delta $ with probability $Q^{n}\left( x+\Delta |x\right) $, accompanying
some SLOCC operations on the quantum output. In the \ following lemma, for a
given $Q^{n}(y|x)$, $\widetilde{Q^{n}}\left( y|x\right) $ is a transition
matrix such that $\widetilde{Q^{n}}\left( y|x\right) =Q^{n}(y|x)$ for $y>x$
and $\widetilde{Q^{n}}\left( y|x\right) =0$ for $y<x$, and $\widetilde{Q^{n}}%
\left( x|x\right) :=Q^{n}(x|x)+\sum_{y<x}Q^{n}(y|x)$.

\begin{lem}
\label{lem:k1} In optimizing (maximizing) (i)-(vi), we can restrict
ourselves to the protocol satisfying (a)-(c).

\begin{itemize}
\item[(i)]  (\ref{g-yield}) under the constraint on the worst-case/average
distortion

\item[(ii)] the weighted sum (\ref{g-y+d})

\item[(iii)] Total fidelity (\ref{def:totalfidelity}).
\end{itemize}

\begin{itemize}
\item[(a)] The protocol is of the form $\left\{ B^{n}\circ C_{\ast
}^{n}\right\} $.

\item[(b)] In $B^{n}$, the corresponding $Q^{n}(y|x)$ satisfies $Q^{n}\left(
y|x\right) =0$ for $y<x$.

\item[(c)] $B^{n}$ does not change quantum output of $C_{\ast }^{n}$.
\end{itemize}
\end{lem}

\begin{pf}
The condition (a) follows from Lemma \ref{optimal}, for worst-case/average
distortion, (\ref{g-yield}), and total fidelity (\ref{def:totalfidelity})
because they satisfy (\ref{inv-measure}) and (\ref{affine-measure}).

For $f(x)$ is monotone increasing,%
\begin{equation*}
\sum_{x,y}f(y)Q^{n}(y|x)Q_{C_{\ast }^{n}}^{\phi }(x)\leq \sum_{x,y}f(y)%
\widetilde{Q^{n}}(y|x)Q_{C_{\ast }^{n}}^{\phi }(x),
\end{equation*}%
where $\widetilde{Q^{n}}\left( y|x\right) $ is a transition matrix such that 
$\widetilde{Q^{n}}\left( y|x\right) =Q^{n}(y|x)$ for $y>x$ and $\widetilde{%
Q^{n}}\left( y|x\right) =0$ for $y<x$, and $\widetilde{Q^{n}}\left(
x|x\right) :=Q^{n}(x|x)+\sum_{y<x}Q^{n}(y|x)$. Hence, $\widetilde{Q^{n}}%
\left( y|x\right) $ improves $Q^{n}(y|x)$ in average yield (\ref{g-yield}),
while worst-case/average distortion is unchanged as is proved later.
Therefore, (b) applies to (i) and (ii).

To go on further, we have to find out optimal state transition made by the
postprocessing. When the postprocessing $B^{n}$changes classical output $x$
to $y$, the corresponding quantum output $\rho _{B^{n}}^{\phi }(y|x)$ which
minimize the distortion i.e., maximizes the fidelity to $\Vert 2^{ny}\rangle 
$ is%
\begin{equation}
\rho _{B^{n}}^{\phi }(y|x)=\left\{ 
\begin{array}{cc}
\Vert 2^{nx}\rangle \langle 2^{nx}|| & (y>x), \\ 
\Vert 2^{ny}\rangle \langle 2^{ny}|| & (y\leq x),%
\end{array}%
\right.   \label{state-opt}
\end{equation}%
for the reasons stated shortly. In case the $y\leq x$, LOCC can change the
output of $C_{\ast }^{n}$, $||2^{nx}\rangle $, to $\Vert 2^{ny}\rangle $
perfectly and deterministically. On the other hand, in case $y>x$,
monotonicity of Schmidt rank by SLOCC implies that $\Vert 2^{nx}\rangle $ is
the best approximate state to $\Vert 2^{ny}\rangle $ in all the states which
can be reached from $\Vert 2^{nx}\rangle $ with non-zero probability. This
transition causes the distortion of $1-2^{-n(y-x)}$.

From (\ref{state-opt}), it is easily understood that worst-case/average
distortion of $\widetilde{Q^{n}}(y|x)$ equals that of $Q^{n}(y|x)$, and that
the condition (c) applies to (i) and (ii).

It remains to prove (b) and (c) for (iii). Observe that total fidelity (\ref%
{def:totalfidelity}) does not depend on the classical output of the
protocol. Therefore, condition (b) is not restriction in optimization.
Therefore, we only prove (c). By definition,%
\begin{equation*}
F_{C^{n}}^{\phi }\left( R\right) =\sum_{x,y}\sum_{z:z\geq
R}Q^{n}(y|x)Q_{C_{\ast }^{n}}^{\phi }(x)\langle 2^{nz}||\rho _{B^{n}}^{\phi
}(y|x)\Vert 2^{nz}\rangle .
\end{equation*}%
In  $x\geq R$ case, $\rho _{B^{n}}^{\phi }(y|x)=\Vert 2^{nx}\rangle \langle
2^{nx}||$ achieves%
\begin{equation*}
\sum_{y}\sum_{z:z\geq R}Q^{n}(y|x)\langle 2^{nz}||\rho _{B^{n}}^{\phi
}(y|x)\Vert 2^{nz}\rangle =1,
\end{equation*}%
which is maximal. In $x<R$ case, for any $z\geq R$($>x$), the maximum of $%
\langle 2^{nz}||\rho _{B^{n}}^{\phi }(y|x)\Vert 2^{nz}\rangle $ is achieved
by $\rho _{B^{n}}^{\phi }(y|x)=\Vert 2^{nx}\rangle \langle 2^{nx}||$,
because monotonicity of Schmidt rank by SLOCC implies that $\Vert
2^{nx}\rangle $ is the best approximate state to $\Vert 2^{nz}\rangle $ in
all the states which can be reached from $\Vert 2^{nx}\rangle $ with
non-zero probability. Therefore, the optimal output state should be as is
described in (c).
\end{pf}

Now, the protocol of interest is very much restricted. We modify classical
output of $C_{\ast }^{n}$ according to transition probability $Q^{n}(y|x)$,
while its quantum output is untouched. Note $Q^{n}(y|x)$ is non-zero only if 
$y\geq x$. Especially, transition to $y$ strictly larger than $x$ means that
the protocol claims the yield $y$ while in fact its yield is $x<y$. In other
words, this is excessive claim on its yield.

Main part of our effort in the following is how to suppress 'excessive
claim', or $Q^{n}(y|x)$ for $y>x$ by setting appropriate measure or
constraint.

Note that the mathematical treatment is much simplified now, for we only
have to optimize transition probability $Q^{n}(y|x)$, with the condition
that the distortion $1-2^{-n(y-x)}$ occurs only if $y\geq x$.

Observe that in the proof of these lemmas, we have used the uncertainty
about Schmidt basis. This assumption is needed to justify the condition (\ref%
{inv-measure}). However, the uncertainty about Schmidt coefficients has
played no role. Therefore, Lemma~\ref{optimal}-\ref{lem:k1} holds true even
in the case where Schmidt coefficients are known.

Hereafter, maximization/minimization over local unitaries will be often
removed, because the protocols of our interest are local unitary invariant.

\subsection{Distortion-free protocols}

\label{subsec:dist-free}

\begin{thm}
$\{C_{\ast }^{n}\}$ achieves the optimal (maximal) value of (\ref{g-yield})
for all universal distortion-free concentrations for all finite $n$, any
input state $\left\vert \phi \right\rangle $, and any threshold $R$. Here, $%
f $ only need to be monotone increasing, and need not to be bounded nor
continuous. \label{opt1}
\end{thm}

\begin{pf}
Lemma~\ref{lem:k1} apply\ to this case, for distortion-free condition
writes, using the invariant measure of a distortion, $\max_{U,V}\epsilon
_{C^{n}}^{U\otimes V\phi }=0$. To increase the value of (\ref{g-yield}), $%
Q^{n}\left( x+\Delta |x\right) $ should be non-zero for some $x$, $\Delta $
with $\Delta >0$, which causes non-zero distortion. Hence, it is impossible
to improve the yield measure (\ref{g-yield}) by postprocessing.
\end{pf}

Observe that the proof also applies to the case where Schmidt coefficients
are known, for the condition $\max_{U,V}\epsilon _{C^{n}}^{U\otimes V\phi
}=0 $ assertion is needed to be imposed only on the input state.

\subsection{Constraints on the worst-case distortion}

\label{subsec:worst-distortion}In this subsection, we discuss the higher
order asymptotic optimality of $\{C_{\ast }^{n}\}$ in terms of the average
yield (\ref{g-yield}) under the constraint on the worst-case distortion, 
\begin{eqnarray*}
&&\max_{U,V}\epsilon _{C^{n}}^{U\otimes V\phi } \\
&=&\max_{\Delta :\exists x,\,Q^{n}(x+\Delta |x)\neq 0}\left( 1-2^{-n\Delta
}\right) \\
&\leq &r_{n}<1,
\end{eqnarray*}%
which implies%
\begin{equation}
Q^{n}(x+\Delta |x)=0,\quad \Delta \geq \frac{-\log (1-r_{n})}{n}.
\label{improve-upper-2}
\end{equation}%
This means that the magnitude of the improvement in the yield is uniformly
upper-bounded by $\frac{-\log (1-r_{n})}{n}$. Ineq. (\ref{improve-upper-2})
is the key to the rest of the argument in this subsection.

In discussing the average yield (\ref{g-yield}), we assume $f(x)$ is
continuously differentiable at around $x=\mathrm{H}\left( \mathbf{p}_{\phi
}\right) $. In addition, first we assume $f^{\prime }\left( \mathrm{H}\left( 
\mathbf{p}_{\phi }\right) \right) >0$. After that, we study the case where $%
f^{\prime }(x)=0$ in the neighborhood of $x=$.$\mathrm{H}\left( \mathbf{p}%
_{\phi }\right) $. A typical example of the former and the latter is $f(x)=%
\frac{x}{\log d}$ and $f(x)=\Theta (x-R)$, respectively.

Note that the argument in this section holds true also for the cases where
Schmidt coefficients are known. This is because the constraint $%
\max_{U,V}\epsilon _{C^{n}}^{U\otimes V\phi }\leq r_{n}$ is needed to be
imposed only on a given input state, and not on all the state.

\begin{thm}
\label{th:opt-g-yield-worst}Suppose that $f$ is continuously differentiable
in a region $(R_{1},R_{2})$ with $\ R_{1}<\mathrm{H}\left( \mathbf{p}_{\phi
}\right) <R_{2}$ , and $r_{n}$ is smaller than $1-\delta $ with $\delta $
being a positive constant, and $r_{n}$ is not exponentially small. Then,

\begin{itemize}
\item[(i)] $\left\{ C_{\ast }^{n}\right\} $ is optimal in the order which is
slightly larger than $O\left( \frac{r_{n}}{n}\right) $, or for any protocol $%
\{C^{n}\}$, 
\begin{equation*}
\mathrm{E}_{Q_{C^{n}}^{\phi }}^{X}f(X)\leq \mathrm{E}_{Q_{C_{\ast
}^{n}}^{\phi }}^{X}f(X)+O\left( \frac{r_{n}}{n}\right) .
\end{equation*}

\item[(ii)] if $f^{\prime }\left( \mathrm{H}\left( \mathbf{p}_{\phi }\right)
\right) >0$, there is a protocol $\{C^{n}\}$ which is better than $\left\{
C_{\ast }^{n}\right\} $ by the magnitude of $O\left( \frac{r_{n}}{n}\right) $
for an input $\left\vert \psi \right\rangle $, or 
\begin{equation*}
\mathrm{E}_{Q_{C^{n}}^{\psi }}^{X}f(X)\geq \mathrm{E}_{Q_{C_{\ast
}^{n}}^{\psi }}^{X}f(X)+O\left( \frac{r_{n}}{n}\right) .
\end{equation*}
\end{itemize}
\end{thm}

Applied to $f(x)=\frac{x}{\log d}$, (i) and (ii) imply that with the
constraint $\max_{U,V}\epsilon _{C^{n}}^{U\otimes V\phi }\rightarrow 0$, $%
\left\{ C_{\ast }^{n}\right\} $ is optimal up to $O\left( \frac{1}{n}\right) 
$-terms, and not optimal in the order smaller than that. Hence, the
coefficients computed in Appendix\thinspace \ref{sec:cal-ave-opt} are
optimal.

\begin{pf}
(i) Obviously, the optimal protocol $\{C^{n}\}$ is given by 
\begin{equation}
Q^{n}(x+\Delta ^{\prime }|x)=1,\quad \Delta ^{\prime }=\left\lfloor \frac{%
-\log (1-r_{n})}{n}\right\rfloor .  \label{opt-worst-dist}
\end{equation}%
In the region $(R_{1},R_{2})$, with $c:=\max_{x:R_{1}\leq x\leq
R_{2}}f^{^{\prime }}\left( x\right) $, 
\begin{eqnarray*}
f\left( x+\Delta ^{\prime }\right) &\leq &f\left( x\right) +c\Delta ^{\prime
} \\
&\leq &f\left( x\right) +c\frac{-\log (1-r_{n})}{n}
\end{eqnarray*}%
holds. For the function $-\log (1-x)$ is monotone and concave, if $%
r_{n}<1-\delta $, we have%
\begin{eqnarray*}
f\left( x+\Delta ^{\prime }\right) &\leq &f\left( x\right) +\frac{cr_{n}}{n}%
\left( \frac{-\log (1-1+\delta )+\log (1-0)}{1-\delta }\right) \\
&\leq &f\left( x\right) +c\frac{-\log (1-\delta )}{1-\delta }\frac{r_{n}}{n}.
\end{eqnarray*}%
The average of the both sides of this over $x$ yields 
\begin{eqnarray*}
\mathrm{E}_{Q_{C^{n}}^{\phi }}^{X}f(X)\leq \mathrm{E}_{Q_{C_{\ast
}^{n}}^{\phi }}^{X}f(X)+c\frac{-\log (1-\delta )}{1-\delta }\frac{r_{n}}{n}%
+O(2^{-nD}) &&, \\
\quad \exists D>0 &&,
\end{eqnarray*}%
for the sum over the complement of $(R_{1},R_{2})$ is exponentially small
due to the third equation of (\ref{grep-type}). This implies the optimality
of our protocol.

(ii) Due to mean value theorem, 
\begin{eqnarray*}
f\left( x+\frac{-\log (1-r_{n})}{n}\right) &>&f\left( x\right) +c^{\prime }%
\frac{-\log (1-r_{n})}{n},\quad \exists c^{\prime }>0, \\
&>&f\left( x\right) +(-c^{\prime }\log \delta )\frac{r_{n}}{n},
\end{eqnarray*}%
holds in a neighborhood of $x=\mathrm{H}\left( \mathbf{p}_{\phi }\right) $.
Hence, letting $\{C^{n}\}$ be the protocol corresponding to (\ref%
{opt-worst-dist}), we have%
\begin{eqnarray*}
\mathrm{E}_{Q_{C^{n}}^{\phi }}^{X}f(X)\geq \mathrm{E}_{Q_{C_{\ast
}^{n}}^{\phi }}^{X}f(X)+(-c^{\prime }\log \delta )\frac{r_{n}}{n}%
-O(2^{-nD}), && \\
\quad \exists D>0, &&
\end{eqnarray*}%
proving the achievability.
\end{pf}

In case of $f(x)=\Theta (x-R)$, which is flat at around $x=\mathrm{H}\left( 
\mathbf{p}_{\phi }\right) $, (ii) of this theorem does not apply, and as is
shown below, the upper-bound to the average yield suggested by (i) is not
tight at all.

\begin{thm}
\label{th:opt-g-yield-worst-2} Suppose $f^{^{\prime }}\left( x\right)
=0\,(R_{1}<x<R_{2})$, $f(R_{1}-0)\neq f\left( \mathrm{H}\left( \mathbf{p}%
_{\phi }\right) \right) $, $f(R_{2}+0)\neq f\left( \mathrm{H}\left( \mathbf{p%
}_{\phi }\right) \right) $, and $\varlimsup_{n\rightarrow \infty }r_{n}<1$.
If $\mathrm{H}\left( \mathbf{p}_{\phi }\right) >R_{1}$,%
\begin{eqnarray*}
&&\varlimsup_{n\rightarrow \infty }\frac{-1}{n}\log \sum_{x:x\leq
R_{1}}\left\{ f\left( \mathrm{H}\left( \mathbf{p}_{\phi }\right) \right)
-f(x)\right\} Q_{C^{n}}^{\phi }\left( x\right) \\
&\leq &\mathrm{D}\left( R_{1}||\mathbf{p}_{\phi }\right) ,
\end{eqnarray*}%
holds. If $\mathrm{H}\left( \mathbf{p}_{\phi }\right) <R_{2}$,%
\begin{eqnarray*}
&&\varliminf_{n\rightarrow \infty }\frac{-1}{n}\log \sum_{x:x\geq
R_{2}}\left\{ f(x)-f\left( \mathrm{H}\left( \mathbf{p}_{\phi }\right)
\right) \right\} Q_{C^{n}}^{\phi }\left( x\right) \\
&\geq &\mathrm{D}\left( R_{2}||\mathbf{p}_{\phi }\right) ,
\end{eqnarray*}%
and the equality is achieved by $\left\{ C_{\ast }^{n}\right\} $.
\end{thm}

This theorem intuitively means that, if $f(x)$ is flat at the neighborhood
of $x=\mathrm{H}\left( \mathbf{p}_{\phi }\right) $, for the optimal
protocol, the quantity (\ref{g-yield}) is approximately of the form, 
\begin{equation*}
f\left( \mathrm{H}\left( \mathbf{p}_{\phi }\right) \right) -A2^{-n\mathrm{D}%
\left( R_{1}||\mathbf{p}_{\phi }\right) }+B2^{-n\mathrm{D}\left( R_{2}||%
\mathbf{p}_{\phi }\right) }.
\end{equation*}%
Applied to $f(x)=\Theta (x-R)$, the theorem implies the optimality of (\ref%
{fexp1-1}) and (\ref{fexp-2}) under the constraint $\varlimsup_{n\rightarrow
\infty }\max_{U,V}\epsilon _{C^{n}}^{U\otimes V\phi }<1$.

\begin{pf}
Suppose $\mathrm{H}\left( \mathbf{p}_{\phi }\right) >R_{1}$ . For any $%
R<R_{1}$, 
\begin{eqnarray*}
&&\sum_{x:x\leq R_{1}}\left\{ f\left( \mathrm{H}\left( \mathbf{p}_{\phi
}\right) \right) -f(x)\right\} Q_{C^{n}}^{\phi }\left( x\right) \\
&\geq &\sum_{x:x\leq R}\left\{ f\left( \mathrm{H}\left( \mathbf{p}_{\phi
}\right) \right) -f(x)\right\} Q_{C^{n}}^{\phi }\left( x\right) \\
&\geq &\left\{ f\left( \mathrm{H}\left( \mathbf{p}_{\phi }\right) \right)
-f(R)\right\} \sum_{x:x\leq R}Q_{C^{n}}^{\phi }\left( x\right) ,
\end{eqnarray*}%
where the second inequality is due to monotonicity of $f$. On the other
hand, (\ref{improve-upper-2}) implies 
\begin{equation*}
\sum_{x:x\leq R}Q_{C^{n}}^{\phi }\left( x\right) \geq \sum_{x:x\leq R-\frac{%
-\log (1-r_{n})}{n}}Q_{C_{\ast }^{n}}^{\phi }\left( x\right) .
\end{equation*}%
Combination of these inequalities with (\ref{fexp1-1}) leads to 
\begin{eqnarray*}
&&\varlimsup_{n\rightarrow \infty }\frac{-1}{n}\log \sum_{x:x\leq
R_{1}}\left\{ f\left( \mathrm{H}(\mathbf{p}_{\phi })\right) -f(x)\right\}
Q_{C^{n}}^{\phi }(x) \\
&\leq &\varlimsup_{n\rightarrow \infty }\frac{-1}{n}\left\{ 
\begin{array}{c}
\log \sum_{x:x\leq R-\frac{-\log (1-r_{n})}{n}}Q_{C_{\ast }^{n}}^{\phi
}\left( x\right) \\ 
+\log \left\{ f\left( \mathrm{H}\left( \mathbf{p}_{\phi }\right) \right)
-f(R)\right\}%
\end{array}%
\right\} \\
&\leq &\mathrm{D}\left( R||\mathbf{p}_{\phi }\right) ,
\end{eqnarray*}%
which, letting $R\rightarrow R_{1}$, leads to the first inequality. On the
other hand, in $\mathrm{H}\left( \mathbf{p}_{\phi }\right) <R_{2}$ case, the
monotonicity of $f$ and (\ref{improve-upper-2}) also implies%
\begin{eqnarray*}
&&\sum_{x:x\geq R_{2}}\left\{ f(x)-f\left( \mathrm{H}\left( \mathbf{p}_{\phi
}\right) \right) \right\} Q_{C^{n}}^{\phi }\left( x\right) \\
&\leq &\left\{ f(\log d)-f\left( \mathrm{H}\left( \mathbf{p}_{\phi }\right)
\right) \right\} \sum_{x:x\geq R_{2}}Q_{C^{n}}^{\phi }\left( x\right) \\
&\leq &\left\{ f(\log d)-f\left( \mathrm{H}\left( \mathbf{p}_{\phi }\right)
\right) \right\} \sum_{x:x\geq R_{2}-\frac{-\log (1-r_{n})}{n}}Q_{C_{\ast
}^{n}}^{\phi }\left( x\right) .
\end{eqnarray*}%
Combination of this with (\ref{fexp-2}) leads to the second inequality. \ 

The achievability is proved as follows. Suppose $\mathrm{H}\left( \mathbf{p}%
_{\phi }\right) >R_{1}$. For $x$ smaller than $R_{1}$,%
\begin{eqnarray*}
f\left( \mathrm{H}\left( \mathbf{p}_{\phi }\right) \right) -f(x) &\leq
&f\left( \mathrm{H}\left( \mathbf{p}_{\phi }\right) \right) \\
&=&f\left( \mathrm{H}\left( \mathbf{p}_{\phi }\right) \right) (1-\Theta
(x-R_{1})).
\end{eqnarray*}%
Hence, the exponent is lower bounded by 
\begin{eqnarray*}
&&\varlimsup_{n\rightarrow \infty }\frac{-1}{n}\log \mathrm{E}_{Q_{C_{\ast
}^{n}}^{\phi }}^{X}\left\{ 1-\Theta (X-R_{1})\right\} \\
&&+\varlimsup_{n\rightarrow \infty }\frac{-1}{n}\log f\left( \mathrm{H}%
\left( \mathbf{p}_{\phi }\right) \right) \\
&=&\mathrm{D}\left( R_{1}||\mathbf{p}_{\phi }\right) ,
\end{eqnarray*}%
which means the first inequality is achieved. Suppose $\mathrm{H}\left( 
\mathbf{p}_{\phi }\right) <R_{2}$. For $x$ larger than $R_{2}$, then we have%
\begin{eqnarray*}
&&f(x)-f\left( \mathrm{H}\left( \mathbf{p}_{\phi }\right) \right) \\
&\geq &\left\{ f(R_{0})-f\left( \mathrm{H}\left( \mathbf{p}_{\phi }\right)
\right) \right\} \Theta (x-R_{0}),
\end{eqnarray*}%
where $R_{0}$ is an arbitrary constant with $R_{0}>R_{2}$. Hence, the
exponent is upper-bounded by%
\begin{eqnarray*}
&&\varlimsup_{n\rightarrow \infty }\frac{-1}{n}\log \mathrm{E}_{Q_{C_{\ast
}^{n}}^{\phi }}^{X}\Theta (x-R_{0}) \\
&&+\varlimsup_{n\rightarrow \infty }\frac{-1}{n}\log \left\{ f\left( \mathrm{%
H}\left( \mathbf{p}_{\phi }\right) \right) -f(R_{0})\right\} \\
&=&\mathrm{D}\left( R_{0}||\mathbf{p}_{\phi }\right) .
\end{eqnarray*}%
Letting $R_{0}\rightarrow R_{1}$, we have the achievability of the second
inequality.
\end{pf}

\subsection{Constraints on the average distortion}

\label{subsec:average-distortion}

In this subsection, we discuss the higher-order asymptotic optimality of $%
\{C_{\ast }^{n}\}$ in terms of the generalized average yield (\ref{g-yield})
under the constraint on the average distortion, 
\begin{eqnarray*}
&&\max_{U,V}\overline{\epsilon }_{C^{n}}^{U\otimes V\phi } \\
&=&\sum_{\Delta }\left( 1-2^{-n\Delta }\right) \mathrm{E}_{Q_{C_{\ast
}^{n}}^{\phi }}^{X}Q^{n}(X+\Delta |X) \\
&\leq &r_{n}<1.
\end{eqnarray*}%
Denote the probability that the improvement by the amount $\Delta $ occurs
by 
\begin{equation*}
\Pr_{\phi }\left( \Delta \right) :=\mathrm{E}_{Q_{C_{\ast }^{n}}^{\phi
}}^{X}Q^{n}(X+\Delta |X).
\end{equation*}%
Observe that 
\begin{eqnarray*}
\overline{\epsilon }_{C^{n}}^{\phi } &=&\sum_{\Delta }\left( 1-2^{-n\Delta
}\right) \Pr_{\phi }\left( \Delta \right) \\
&\geq &\left( 1-2^{-c}\right) \Pr_{\phi }\left\{ \Delta \geq \frac{c}{n}%
\right\}
\end{eqnarray*}%
which implies 
\begin{equation}
\Pr_{\phi }\left\{ \Delta \geq \frac{c}{n}\right\} \leq \frac{r_{n}}{1-2^{-c}%
}.  \label{improve-upper}
\end{equation}%
Hence, the magnitude of improvement is upper-bounded only in average sense,
in contrast with (\ref{improve-upper-2}) which implies an upper-bound
uniform with respect to $x$.

Suppose that $f$ is continuously differentiable all over the region $(0,\log
d)$. Then, the improvement $x\rightarrow x+\Delta $ causes the distortion by
the amount of 
\begin{eqnarray*}
1-2^{-n\Delta } &\geq &\frac{1-d^{-n}}{\log d}\Delta \\
&\geq &\frac{1-d^{-n}}{\log d}\frac{1}{c}\left( f\left( x\right) -f\left(
x-\Delta \right) \right) ,
\end{eqnarray*}%
where, $c=\max_{x:0\leq x\leq 1}f^{\prime }\left( x\right) .$ Taking average
of the both side, 
\begin{equation*}
\overline{\epsilon }_{C^{n}}^{\phi }\geq \frac{1-d^{-n}}{\log d}\frac{1}{c}%
\left( \mathrm{E}_{Q_{C^{n}}^{\phi }}^{X}f\left( X\right) -\mathrm{E}%
_{Q_{C_{\ast }^{n}}^{\phi }}^{X}f\left( X\right) \right) ,
\end{equation*}%
or,%
\begin{equation*}
\mathrm{E}_{Q_{C^{n}}^{\phi }}^{X}f\left( X\right) \leq \mathrm{E}%
_{Q_{C_{\ast }^{n}}^{\phi }}^{X}f\left( X\right) +\frac{c\log d}{1-d^{-n}}%
\overline{\epsilon }_{C^{n}}^{\phi }.
\end{equation*}

On the other hand, let the protocol $\{C^{n}\}$ be the one corresponding to 
\begin{equation*}
Q^{n}\left( \log d|x\right) =r_{n},\quad \forall x.
\end{equation*}%
Then, we have%
\begin{eqnarray*}
\mathrm{E}_{Q_{C^{n}}^{\phi }}^{X}f\left( X\right) &=&\mathrm{E}_{Q_{C_{\ast
}^{n}}^{\phi }}^{X}f\left( X\right) +r_{n}\mathrm{E}_{Q_{C_{\ast
}^{n}}^{\phi }}^{X}\left\{ 1-f(X)\right\} \\
&\geq &\mathrm{E}_{Q_{C_{\ast }^{n}}^{\phi }}^{X}f\left( X\right) \\
&&+r_{n}\left\{ 1-f(\mathrm{H}(\mathbf{p}_{\phi })+c)\right\} \left( 1-2^{-n%
\mathrm{D}(H(\mathbf{p}_{\phi })+c||\mathbf{p}_{\phi })}\right) , \\
&&\quad \quad \quad \quad \quad \quad \quad \quad \quad \quad \quad \quad
\quad \quad \quad \quad \quad \quad \quad \forall c>0,
\end{eqnarray*}%
while the average distortion of $\{C^{n}\}$ is at most $r_{n}$.

Now, we extend these arguments to the case where finitely many discontinuous
points exist. First, in the proof of the upper-bound, it is sufficient for $%
f $ to be continuously differentiable in the neighborhood of $x=\mathrm{H}%
\left( \mathbf{p}_{\phi }\right) $, if the exponentially small terms are
neglected. Second, the evaluation of the performance of the protocol
constructed above does not rely on the differentiability of $f$. Therefore,
we have the following theorem.

\begin{thm}
\label{opt-g-yield}

\begin{itemize}
\item[(i)] Suppose that $f$ is continuously differentiable in the
neighborhood of $x=\mathrm{H}\left( \mathbf{p}_{\phi }\right) $. Suppose
also $r_{n}$ is not exponentially small. Then, if $\overline{\epsilon }%
_{C^{n}}^{\phi }\leq r_{n}$., $\{C_{\ast }^{n}\}$ is optimal in terms of (%
\ref{g-yield}), up to the order which is slightly larger than $O(r_{n})$, or
for any protocol $\{C_{n}\}$, 
\begin{equation*}
\mathrm{E}_{Q_{C^{n}}^{\phi }}^{X}f\left( X\right) \leq \mathrm{E}%
_{Q_{C_{\ast }^{n}}^{\phi }}^{X}f\left( X\right) +O(r_{n}).
\end{equation*}

\item[(ii)] Suppose that $f(\mathrm{H}(\mathbf{p}_{\phi })+c)<1$, $\exists
c>0$, and $\overline{\epsilon }_{C^{n}}^{\phi }\leq r_{n}$. Then, there is a
protocol $\{C^{n}\}$ which improves $\{C_{\ast }^{n}\}$ by the order of $%
O(r_{n})$;%
\begin{equation*}
\mathrm{E}_{Q_{C^{n}}^{\phi }}^{X}f\left( X\right) \geq \mathrm{E}%
_{Q_{C_{\ast }^{n}}^{\phi }}^{X}f\left( X\right) +O(r_{n}).
\end{equation*}
\end{itemize}
\end{thm}

Let us compare this theorem with Theorem~\ref{th:opt-g-yield-worst} which
states optimality results with constraint on worst-case distortion. First,
the yield is worse by the order of $\frac{1}{n}$. In particular, if $f(x)=%
\frac{x}{\log d}$, $r_{n}$ needs to be $o(\frac{\log n}{n})$ for optimality
up to a higher order term is guaranteed. By contrast, under the constraint
on the worst-case distortion, $r_{n}=o(1)$ is enough to certify optimality
up to the third leading term.

Second, applied to the case of $f(x)=\Theta (x-R)$, $\mathrm{H}(\mathbf{p}%
_{\phi })>R$ , Theorem~\ref{opt-g-yield} implies the following. With $%
r_{n}=o(1)$, the success probability $\sum_{x:x\geq R}Q_{C^{n}}^{\phi
}\left( x\right) $ vanishes (strong converse holds), but the speed of
convergence is at most as fast as $r_{n}$, which is not exponentially fast,
in general. Therefore, (\ref{fexp-2}) is far from optimal unless $r_{n}$
decreases exponentially fast. By contrast, under the constraint on the
worst-case distortion, a constant upper-bound is enough to guarantee the
optimality of the exponent (\ref{fexp-2}).

Let us study the equivalence of Theorem~\ref{th:opt-g-yield-worst-2},
because Theorem~\ref{opt-g-yield}, (ii) cannot be applied to discussion of
optimality of the exponent~(\ref{fexp1-1}), in which the rate $R$ is
typically less than $\mathrm{H}(\mathbf{p}_{\phi })$.

\begin{lem}
\label{lem:improve-upper}Suppose that%
\begin{equation*}
r\geq \varlimsup_{n\rightarrow \infty }\max_{U,V}\overline{\epsilon }%
_{C^{n}}^{U\otimes V\phi }
\end{equation*}%
holds for all $\left\vert \phi \right\rangle $. Then, for all $\left\vert
\phi \right\rangle $, all $c>0$, all $\delta >0$, and all $R^{\prime }$, $%
R^{\prime \prime }$ with $R^{\prime }>R^{\prime \prime }$, there is a
sequence $\{x_{n}\}$ such that $R^{\prime \prime }\leq x_{n}\leq R^{\prime }$
and 
\begin{equation*}
\varlimsup_{n\rightarrow \infty }\sum_{\Delta :\Delta \geq \frac{c}{n}%
}Q^{n}(x_{n}+\Delta |x_{n})\leq \frac{r+\delta }{1-2^{-c}}
\end{equation*}%
hold.
\end{lem}

\begin{pf}
Assume the lemma is false, i.e., there is a sequence $\{n_{k}\}$ such that
for all $x$ in the interval $(R^{\prime \prime },R^{\prime })$, 
\begin{equation*}
\sum_{\Delta :\Delta \geq c}Q^{n_{k}}(x+\Delta |x)>\frac{r+\delta }{1-2^{-c}}
\end{equation*}%
holds. Choosing $\left\vert \phi \right\rangle $ with $R^{\prime }>\mathrm{H}%
\left( \boldsymbol{p}_{\phi }\right) >R^{\prime \prime }$, we have%
\begin{eqnarray*}
&&\Pr_{\phi }\left\{ \Delta \geq \frac{c}{n_{k}}\right\} \\
&\geq &\sum\limits_{x:R^{\prime \prime }\leq x\leq R^{\prime }}\sum_{\Delta
:\Delta \geq \frac{c}{n_{k}}}Q^{n_{k}}(x+\Delta |x)Q_{C_{\ast
}^{n_{k}}}^{\phi }(x) \\
&\geq &\frac{r+\delta }{1-2^{-c}}\left( 1-2^{-n_{k}\left( \min \left\{ 
\mathrm{D}(R^{\prime }||\mathbf{p}_{\phi }),\mathrm{D}(R^{\prime }||\mathbf{p%
}_{\phi })\right\} -\delta ^{\prime }\right) }\right) , \\
&&\quad \quad \quad \quad \quad \quad \quad \quad \quad \quad \forall \delta
^{\prime }>0,\,\exists k_{1},\,\,\forall k\geq k_{1},
\end{eqnarray*}%
which, combined with (\ref{improve-upper}), implies 
\begin{equation*}
\frac{r+\delta }{1-2^{-c}}\left( 1-2^{-n_{k}\left( \min \left\{ \mathrm{D}%
(R^{\prime }||\mathbf{p}_{\phi }),\mathrm{D}(R^{\prime \prime }||\mathbf{p}%
_{\phi })\right\} -\delta ^{\prime }\right) }\right) \leq \frac{r}{1-2^{-c}}.
\end{equation*}%
This cannot hold when $n_{k}$ is large enough. Therefore, the lemma has to
be true.
\end{pf}

\begin{thm}
\label{th:opt-exponent}Suppose that $\ f(x)=1$ for $x\geq R$, and $\mathrm{H}%
(\mathbf{p}_{\phi })>R$ holds. Then, if $\varlimsup_{n\rightarrow \infty
}\max_{U,V}\overline{\epsilon }_{C^{n}}^{U\otimes V\phi }<1$ holds for all $%
\left\vert \phi \right\rangle $, 
\begin{equation}
\varlimsup_{n\rightarrow \infty }\frac{-1}{n}\log \mathrm{E}%
_{Q_{C^{n}}^{\phi }}^{X}\left\{ 1-f\left( X\right) \right\} \leq D(R||%
\mathbf{p}_{\phi }),  \label{g-fexp}
\end{equation}%
and the equality is achieved by $\left\{ C_{\ast }^{n}\right\} $.
\end{thm}

Note the premise of the theorem is the negation of the premise of Theorem~%
\ref{opt-g-yield}\ , (ii). Note also that the constraint on the average
distortion is very moderate, allowing constant distortion.

\begin{pf}
First, we prove (\ref{g-fexp}) for $f(x)=\Theta \left( x-R\right) $. Without
loss of generality, we can assume that $Q^{n}\left( y|x\right) $ is non-zero
only if $y=R$ and $x<R$. Therefore%
\begin{eqnarray*}
&&1-\mathrm{E}_{Q_{C^{n}}^{\phi }}^{X}\left\{ f\left( X\right) \right\} \\
&=&1-\sum\limits_{x<R}Q^{n}\left( R|x\right) Q_{C_{\ast }^{n}}^{\phi }\left(
x\right) \\
&=&1-\sum\limits_{x}Q^{n}\left( R|x\right) Q_{C_{\ast }^{n}}^{\phi }\left(
x\right) \\
&=&\sum\limits_{x}\left( 1-Q^{n}\left( R|x\right) \right) Q_{C_{\ast
}^{n}}^{\phi }\left( x\right) .
\end{eqnarray*}%
Let $R^{\prime }$, $R^{\prime \prime }$ be real numbers with $R^{\prime
}<R^{\prime \prime }<R$, and $\left\{ x_{n}\right\} $ be a sequence given by
Lemma~\ref{lem:improve-upper}. Then we have, due to Lemma~\ref%
{lem:improve-upper},%
\begin{eqnarray*}
&&1-\mathrm{E}_{Q_{C^{n}}^{\phi }}^{X}\left\{ f\left( X\right) \right\} \\
&\geq &\left\{ 1-Q^{n}\left( R|x_{n}\right) \right\} Q_{C_{\ast }^{n}}^{\phi
}\left( x_{n}\right) \\
&\geq &\left( 1-\frac{r+\delta }{1-2^{-n(R-R^{\prime })}}\right) Q_{C_{\ast
}^{n}}^{\phi }\left( x_{n}\right) , \\
&&\quad \quad \quad \quad \quad \quad \quad \quad \quad \quad \quad \quad
\exists n_{0}\,\forall n>n_{0},
\end{eqnarray*}%
which implies 
\begin{eqnarray*}
&&\varlimsup_{n\rightarrow \infty }\frac{-1}{n}\log \sum_{x}\left\{ 1-\Theta
\left( x-R\right) \right\} Q_{C^{n}}^{\phi }(x) \\
&\leq &\lim_{n\rightarrow \infty }\frac{-1}{n}\left\{ 
\begin{array}{c}
\log Q_{C_{\ast }^{n}}^{\phi }\left( R^{\prime \prime }\right) \\ 
+\log \left( 1-\frac{r+\delta }{1-2^{-n(R-R^{\prime })}}\right)%
\end{array}%
\right\} \\
&=&\mathrm{D}\left( R^{\prime \prime }||\mathbf{p}_{\phi }\right) .
\end{eqnarray*}%
For this holds true for all $R^{\prime \prime }<R^{\prime }<R,$ the limit $%
R^{\prime \prime }\rightarrow R$ leads to the inequality (\ref{g-fexp}) for $%
f(x)=\Theta \left( x-R\right) $.

As for $f$ which satisfies the premise of the theorem, we lower-bound $%
1-f(x) $ by $(1-f(R_{0}))(1-\Theta (x-R_{0}))$, with $R_{0}<R$. Then, the
exponent is 
\begin{eqnarray*}
&&\varlimsup_{n\rightarrow \infty }\frac{-1}{n}\log \mathrm{E}%
_{Q_{C^{n}}^{\phi }}^{X}\left\{ 1-f\left( X\right) \right\} \\
&\leq &\varlimsup_{n\rightarrow \infty }\frac{-1}{n}\log \mathrm{E}%
_{Q_{C^{n}}^{\phi }}^{X}(1-\Theta (X-R_{0})) \\
&&+\varlimsup_{n\rightarrow \infty }\frac{-1}{n}\log (1-f(R_{0})) \\
&=&\mathrm{D}\left( R_{0}||\mathbf{p}_{\phi }\right) .
\end{eqnarray*}%
The limit $R_{0}\rightarrow R$ leads to the inequality of the theorem. The
achievability is proven in the same way as the proof of Theorem~\ref%
{th:opt-g-yield-worst}.
\end{pf}

Note the arguments in the proof of Theorem~\ref{opt-g-yield} apply also to
the case where Schmidt coefficients are known. This is because the
constraint $\max_{U,V}\overline{\epsilon }_{C^{n}}^{U\otimes V\phi }\leq
r_{n}$ is required only for a given input state, and not for all the state.
On the contrary, the proof of Theorem~\ref{th:opt-exponent} is valid only if
the constraint $\max_{U,V}\overline{\epsilon }_{C^{n}}^{U\otimes V\phi }\leq
r_{n}$ is assumed for all $\left\vert \phi \right\rangle $ (otherwise, Lemma~%
\ref{lem:improve-upper} cannot be proved), meaning the generalization to the
case where Schmidt coefficients are known is impossible.

\subsection{Weighted sum of the distortion and the yield}

\label{subsec:weighted}In this subsection, we discuss the weighted sum (\ref%
{g-y+d}) of the distortion and the yield. First, we study the case where $f$
is continuously differentiable over the domain, and then study the case
where $f$ is an arbitrary monotone non-decreasing, bounded function. In this
section, we prove (a sort of) non-asymptotic optimality. Finally, we apply
our result to induce another proof of Theorem~\ref{opt-g-yield}. The
argument in this subsection generalize to the case where Schmidt
coefficients are known, as is explained toward the end of Subsection~\ref%
{subsec:whatisproved}.

To have a reasonable result, the wight $\lambda $ cannot be too small nor
too large. In this subsection, we assume 
\begin{equation}
\lambda >1,  \label{weight}
\end{equation}%
because otherwise the yield $f\left( x\right) $ can take the value larger
than the maximum value of the distortion, which equals unity. No explicit
upper-bound to $\lambda $ is assumed, but $\lambda $ is regarded as a
constant only slightly larger than $1$.

Due to Lemma~\ref{lem:k1}, the difference between the value of the measure (%
\ref{g-y+d}) of $\left\{ C_{\ast }^{n}\right\} $ and the protocol
characterized by $Q^{n}\left( x|y\right) $ is

\begin{eqnarray}
&&\sum\limits_{\substack{ x,y:  \\ x\geq y}}\left\{ f\left( x\right)
-f\left( y\right) -\lambda \left( 1-2^{-n\left( x-y\right) }\right) \right\}
\notag \\
&&\times Q^{n}\left( x|y\right) Q_{C_{\ast }^{n}}^{\phi }(x).  \label{diff}
\end{eqnarray}%
For $\left\{ C_{\ast }^{n}\right\} $, or $Q^{n}\left( x|y\right) =0$ to be
optimal, the coefficient for $Q^{n}\left( x|y\right) $ has to be
non-positive, or,%
\begin{equation*}
f\left( x\right) -f\left( y\right) \leq \lambda \left( 1-2^{-n\left(
x-y\right) }\right) .
\end{equation*}%
When $n$ is large enough, the RHS of this approximately equals $\lambda
\Theta \left( x-y\right) $. Hence, this inequality holds, if $n$ is larger
than some threshold $n_{0}$, for varieties of $f$'s. More rigorously, the
condition for the optimality of $\left\{ C_{\ast }^{n}\right\} $ writes 
\begin{eqnarray}
n\geq \frac{-1}{x-y}\log \left( 1-\frac{f\left( x\right) -f(y)}{\lambda }%
\right) , &&  \label{cond-n-delta} \\
\forall x,y,\quad 0\leq y<x\leq \log d. &&  \notag
\end{eqnarray}%
Observe that $-\log (1-x)$ is convex and monotone increasing, and $-\log
(1-0)=0$. Hence, the RHS of (\ref{cond-n-delta}) is upper-bounded by 
\begin{eqnarray*}
&&\left\{ -\log \left( 1-\frac{f\left( \log d\right) -f(0)}{\lambda }\right)
+\log \left( 1-\frac{0}{\lambda }\right) \right\} \\
&&\times \frac{1}{f\left( \log d\right) -f(0)}\frac{f(x)-f(y)}{x-y}, \\
&=&-\log \left( 1-\frac{1}{\lambda }\right) \frac{f(x)-f(y)}{x-y}.
\end{eqnarray*}%
If the function $f$ is continuously differentiable, the last side of the
equation equals%
\begin{equation*}
-\log \left( 1-\frac{1}{\lambda }\right) \max_{x:0\leq x\leq \log
d}f^{\prime }(x)
\end{equation*}%
After all, we have the following theorem.

\begin{thm}
If $f$ satisfies (\ref{cond-f})-(\ref{cond-f-3}), $\left\{ C_{\ast
}^{n}\right\} $ is optimal, i,e., achieves maximum of (\ref{g-y+d}) with the
weight (\ref{weight}) for any input state, any $n$ larger than the threshold 
$n_{0}$, where%
\begin{equation*}
n_{0}=-\log \left( 1-\frac{1}{\lambda }\right) \max_{x:0\leq x\leq \log
d}f^{\prime }(x).
\end{equation*}%
.\label{weitht-semi-asymptotic}
\end{thm}

Some comments on the theorem are in order. First,this assertion is different
from so called asymptotic optimality, in which the higher order terms are
neglected. On the contrary, our assertion is more like non-asymptotic
arguments, for we have proved that $\left\{ C_{\ast }^{n}\right\} $ is
optimal up to arbitrary order if $n$ is larger than some finite threshold.

Second, the factor $-\log \left( 1-\frac{1}{\lambda }\right) $ is relatively
small even if $\lambda $ is very close to $1$. For example, for $\lambda
=1.001$, $-\log \left( 1-\frac{1}{\lambda }\right) =9.96\cdots $. Hence, the
threshold value is not so large. For example, if $f(x)=x$, $n_{0}=\frac{-1}{%
\log d}\log \left( 1-\frac{1}{\lambda }\right) =\frac{9.96\cdots }{\log d}%
\leq 10$.

So far, $\lambda $ has been a constant, but, let $\lambda =\frac{1}{1-d^{-n}}
$, so that the range of $f$ and the distortion coincide with each other.
(Note $\frac{1}{1-d^{-n}}$ is only slightly larger than $1$.) Then, if $f(x)=%
\frac{x}{\log d}$, the condition for the optimality of $\left\{ C_{\ast
}^{n}\right\} $ writes, 
\begin{equation*}
n\geq \frac{-\log \left( 1-\frac{1}{1-d^{-n}}\right) }{\log d}=n
\end{equation*}%
which holds for all $n\geq 1$, implying the non-asymptotic optimality.

Now, let us study the case where $f$ is not differentiable, such as $%
f(x)=\Theta (x-R)$. In such case, we see the condition (\ref{cond-n-delta})
as a restriction on $\Delta :=x-y$ such that $Q^{n}\left( y+\Delta |y\right) 
$ for the optimal protocol takes non-zero value for some $y$. Observe that $%
f(x)-f(y)\leq 1$ is true for all the function satisfying (\ref{cond-f-2}).
Therefore, for 
\begin{equation*}
\Delta \geq \frac{-\log \left( 1-\frac{1}{\lambda }\right) }{n}.
\end{equation*}%
$Q^{n}\left( y+\Delta |y\right) $'s for the optimal protocol vanish for all $%
y$. Hence, the improvement of $\left\{ C_{\ast }^{n}\right\} $ is possible
only in the very small range of $\Delta $, when $n$ is very large.

Those analysis of the weighted sum measures can be applied to the proof of
Theorem~\ref{opt-g-yield}, (i). For this purpose, we use a Lagrangian such
that 
\begin{equation*}
\mathfrak{L}\mathfrak{:}=\min_{U,V}\mathrm{E}_{Q_{C^{n}}^{U\otimes V\phi
}}^{X}f(X)+\lambda \max_{U,V}\left( r_{n}-\overline{\epsilon }%
_{C^{n}}^{U\otimes V\phi }\right) ,
\end{equation*}%
with $\lambda >1$, $r_{n}>0$. Observe that (\ref{g-yield}) cannot be larger
than $\mathfrak{L}$ under the condition $\max_{U,V}\overline{\epsilon }%
_{C^{n}}^{U\otimes V\phi }\leq r_{n}$. For under this condition, the second
term is positive, and the first term of $\mathfrak{L}$ is nothing but (\ref%
{g-yield}). In case that $f$ is differentiable, Theorem~\ref%
{weitht-semi-asymptotic} implies that the maximum of $\mathfrak{L}$ ($%
\lambda >1$) is achieved by $\{C_{\ast }^{n}\}$. Therefore, for $\overline{%
\epsilon }_{C_{\ast }^{n}}^{\phi }=0$, we obtain the inequality,%
\begin{equation}
\max \mbox{(\ref{g-yield})}\leq \mbox{(\ref{g-yield}) for ${C_*^n}$}+\lambda
r_{n}.  \label{upper-g-yield}
\end{equation}%
The similar argument applies to the case where $f$ is continuously
differentiable only in the neighbor of $x=\mathrm{H}(\mathbf{p}_{\phi })$,
except for the exponentially small terms.

\subsection{Total fidelity $F_{C^{n}}^{\protect\phi }\left( R\right) $}

\label{subsec:totalfidelity} This measure equals fidelity between an output
and a target, and relation between $R$ and this quantity reflects the
trade-off between yield and distortion. Note that the argument in this
subsection also generalizes to the case where the Schmidt coefficients are
known, as is explained toward the end of Subsection~\ref{subsec:whatisproved}%
.

\begin{thm}
$\{C_{\ast }^{n}\}$ achieves the optimal (maximum) value of total fidelity (%
\ref{def:totalfidelity}) in all the protocols, for any $n$, any input state $%
\left\vert \phi \right\rangle $, and any threshold $R$.
\end{thm}

\begin{pf}
Due to Lemma \ref{lem:k1}, the protocol of interest is postprocessing of $%
\{C_{\ast }^{n}\}$ which does not touch its quantum output. The total
fidelity of such protocol equals that of $\{C_{\ast }^{n}\}$, for total
fidelity (\ref{def:totalfidelity}) is not related to a classical output of
the protocol.
\end{pf}

\section{Universal concentration as an estimate of entanglement}

\label{sec:estimate} In this section, universal concentration is related to
statistical estimation of entanglement measure $\mathrm{H}(\mathbf{p}_{\phi
})$.

Observe that 
\begin{equation*}
\mathrm{\hat{H}}_{C^{n}}:=\frac{1}{n}\log (\mbox{dim. of max. ent.})
\end{equation*}%
is a natural estimate of $\mathrm{H}(\mathbf{p}_{\phi })$ when $n\gg 1$.
Letting $f(x)=\Theta (x-R)$, Theorem~\ref{opt-g-yield}, (i) implies that the
probability for $\mathrm{\hat{H}}_{C^{n}}>\mathrm{H}(\mathbf{p}_{\phi })$
tends to vanish, if $\varlimsup_{n\rightarrow \infty }\overline{\epsilon }%
_{C^{n}}^{\phi }<1$, as is demonstrated right after the statement of the
theorem. Therefore, if $\{C^{n}\}$ achieves the entropy rate, the estimate $%
\mathrm{\hat{H}}_{C^{n}}$ converges to $\mathrm{H}(\mathbf{p}_{\phi })$ in
probability as $n\rightarrow \infty $ (a \textit{consistent estimate}).
Especially, for the estimate $\widehat{\mathrm{H}}_{C_{\ast }^{n}}$ which is
based on $\left\{ C_{\ast }^{n}\right\} $, the error exponent is given using
(\ref{fexp1-1}) and (\ref{fexp-2}) as, 
\begin{eqnarray*}
&&\varlimsup_{n\rightarrow \infty }\frac{-1}{n}\log \max_{U,V}\mathrm{%
\Pr_{U\otimes V\phi }}\left\{ \left\vert \widehat{\mathrm{H}}_{C_{\ast
}^{n}}-\mathrm{H}(\mathbf{p}_{\phi })\right\vert <\delta \right\} \\
&=&\min_{\left\vert \mathrm{H}(\mathbf{q})-\mathrm{H}(\mathbf{p}_{\phi
})\right\vert \geq \delta }\mathrm{D}(\mathbf{q}\Vert \mathbf{p}_{\phi })
\end{eqnarray*}%
Now, we prove that this exponent is better than any other consistent
estimate which potentially uses global measurements, if the Schmidt basis is
unknown.

\begin{thm}
\begin{eqnarray}
&&\varlimsup_{n\rightarrow \infty }\frac{-1}{n}\log \max_{U,V}\mathrm{%
\Pr_{U\otimes V\phi }}\left\{ \widehat{\mathrm{H}}_{n}<\mathrm{H}(\mathbf{p}%
_{\phi })-\delta \right\}  \notag \\
&\leq &\min_{\mathrm{H}(\mathbf{q})\leq \mathrm{H}(\mathbf{p}_{\phi
})-\delta }\mathrm{D}(\mathbf{q}\Vert \mathbf{p}_{\phi })  \label{est-exp}
\end{eqnarray}

\begin{eqnarray}
&&\varlimsup_{n\rightarrow \infty }\frac{-1}{n}\log \max_{U,V}\mathrm{%
\Pr_{U\otimes V\phi }}\left\{ \widehat{\mathrm{H}}_{n}>\mathrm{H}(\mathbf{p}%
_{\phi })+\delta \right\}  \notag \\
&\leq &\min_{\mathrm{H}(\mathbf{q})\geq \mathrm{H}(\mathbf{p}_{\phi
})+\delta }\mathrm{D}(\mathbf{q}\Vert \mathbf{p}_{\phi })  \label{est-exp2}
\end{eqnarray}%
holds for any consistent estimate $\mathrm{\hat{H}}_{n}$ of $\mathrm{H}(%
\mathbf{p}_{\phi })$ by global measurement, if the Schmidt basis is unknown. %
\label{theorem:bahadur}
\end{thm}

\begin{pf}
An argument almost parallel to the one in Subsection~\ref{subsec:keylemmas}
implies that we can restrict ourselves to the estimate which is computed
from the classical output of $C_{\ast }^{n}$.

From here, we use the argument almost parallel with the one in ~\cite%
{Nagaoka}. For $\epsilon $, $|\phi \rangle $, $|\psi \rangle $ with $\mathrm{%
H}(\mathbf{p}_{\psi })<\mathrm{H}(\mathbf{p}_{\phi })-\delta $, consistency
of $\widehat{\mathrm{H}}_{n}$ implies 
\begin{align}
\mathrm{Pr}_{\phi }\left\{ \widehat{\mathrm{H}}_{n}<\mathrm{H}(\mathbf{p}%
_{\phi })-\delta \right\} (& :=p_{n})\rightarrow 0,  \notag \\
\mathrm{Pr}_{\psi }\left\{ \widehat{\mathrm{H}}_{n}<\mathrm{H}(\mathbf{p}%
_{\phi })-\delta \right\} (& :=q_{n})\rightarrow 1.  \label{converse2}
\end{align}%
On the other hand, monotonicity of relative entropy implies 
\begin{align*}
& \mathrm{D}(Q_{C_{\ast }^{n}}^{\psi }\Vert Q_{C_{\ast }^{n}}^{\phi })\geq 
\mathrm{D}(\mathrm{Pr}_{\psi }\{\mathrm{\hat{H}}_{n}\}\Vert \mathrm{Pr}%
_{\phi }\{\mathrm{\hat{H}}_{n}\}) \\
& \geq q_{n}\log \frac{q_{n}}{p_{n}}+(1-q_{n})\log \frac{1-q_{n}}{1-p_{n}},
\end{align*}%
or, equivalently, 
\begin{align*}
& \frac{-1}{n}\log {p_{n}} \\
& \leq \frac{1}{nq_{n}}\left( \mathrm{D}(Q_{C_{\ast }^{n}}^{\psi }\Vert
Q_{C_{\ast }^{n}}^{\phi })+\mathrm{h}(q_{n})+(1-q_{n})\log (1-p_{n})\right) ,
\end{align*}%
with $\mathrm{h}(x):=-x\log x-(1-x)\log (1-x)$. With the help of Eqs.~(\ref%
{converse2}), letting $n\rightarrow \infty $ of the both sides of this
inequality,.we obtain Bahadur-type inequality~\cite{Bahadur}, 
\begin{equation}
\mbox{ LHS of ~(\ref{est-exp})}\leq \varlimsup_{n\rightarrow \infty }\frac{1%
}{n}\mathrm{D}(Q_{C_{\ast }^{n}}^{\psi }\Vert Q_{C_{\ast }^{n}}^{\phi }),
\label{bahadur2}
\end{equation}%
whose RHS equals $\mathrm{D}(\mathbf{p}_{\psi }\Vert \mathbf{p}_{\phi })$,
as in Appendix~\ref{sec:cal-divergence}. Therefore, choosing $|\psi \rangle $
such that $H(\mathbf{p}_{\psi })$ is infinitely close to $R$, (\ref{est-exp}%
) is proved. (\ref{est-exp2}) is proved almost in the same way.
\end{pf}

\begin{pf}
\textbf{of (\ref{g-fexp}) with }$f(x)=\Theta \left( x-R\right) $, $R>\mathrm{%
H}(\mathbf{p}_{\phi })$

If a protocol $\{C^{n}\}$ satisfies $\varlimsup_{n\rightarrow \infty }%
\overline{\epsilon }_{C^{n}}^{\phi }<1$ and achieves the rate of the entropy
of entanglement, as is mentioned at the beginning of this section, the
corresponding estimate $\mathrm{\hat{H}}_{C^{n}}$ is consistent, and
satisfies Ineq.~(\ref{est-exp}). This is equivalent to the optimality (\ref%
{g-fexp}) with\textbf{\ }$f(x)=\Theta \left( x-R\right) $, $R>\mathrm{H}(%
\mathbf{p}_{\phi })$, for the error probability of $\mathrm{\hat{H}}_{C^{n}}$
equals that of $C^{n}$.
\end{pf}

Suppose in addition that the Schmidt basis is known, and we discuss the
first main term of the mean square error, 
\begin{equation*}
\mathrm{E}_{\phi }(\mathrm{\hat{H}}_{n}-\mathrm{H}(\mathbf{p}_{\phi }))^{2}=%
\frac{1}{n}\tilde{\mathrm{V}}_{\phi }+o\left( \frac{1}{n}\right)
\end{equation*}%
or, 
\begin{equation*}
\tilde{\mathrm{V}}_{\phi }:=\varliminf_{n\rightarrow \infty }n\mathrm{E}%
_{\phi }(\mathrm{\hat{H}}_{n}-\mathrm{H}(\mathbf{p}_{\phi }))^{2}.
\end{equation*}

\begin{thm}
Suppose the Schmidt basis of $|\phi \rangle $ is known and its Schmidt
coefficient is unknown. Then, any global measurement satisfies, 
\begin{equation}
\tilde{\mathrm{V}}_{\phi }\geq \sum_{i=1}^{d}p_{\phi ,i}(\log p_{\phi ,i}-%
\mathrm{H}(\mathbf{p}_{\phi }))^{2},  \label{bound}
\end{equation}%
if $\mathrm{E}_{\phi }(\mathrm{\hat{H}}_{n})\rightarrow \mathrm{H}(\mathbf{p}%
_{\phi })$ ($n\rightarrow \infty $) for all $\left\vert \phi \right\rangle $%
, and the estimate $\mathrm{\hat{H}}_{C_{\ast }^{n}}$ based on $\{C_{\ast
}^{n}\}$ achieves the equality. \label{theorem:meansquare}
\end{thm}

\begin{pf}
Consider a family of state vectors $\left\{ \sum\limits_{i}\sqrt{p_{\phi ,i}}%
|e_{j,A}\rangle |e_{j,B}\rangle \right\} $, where $\{|e_{j,A}\rangle
|e_{j,B}\rangle \}$ is fixed and $\mathbf{p_{\phi }}$ runs over all the
probability distributions supported on $\{1,\ldots ,d\}$. Due to Theorem~5
in \cite{Matsumoto1}, the asymptotically optimal estimate of $\mathrm{H}(%
\mathbf{p}_{\phi })$ is a function of data result from the projection
measurement $\{|e_{j,A}\rangle |e_{j,B}\rangle \}$ on each copies.
Therefore, the problem reduces to the optimal estimate of $\mathrm{H}(%
\mathbf{p}_{\phi })$ from the data generated from probability distribution $%
\mathbf{p}_{\phi }$.

Due to asymptotic Cram{\'{e}}r-Rao inequality of classical statistics, the
asymptotic mean square error of such estimate is lower-bounded by,%
\begin{equation*}
\frac{1}{n}\sum_{1\leq i,j\leq d-1}\left( J^{-1}\right) ^{i,j}\frac{\partial 
\mathrm{H}}{\partial p_{i}}\frac{\partial \mathrm{H}}{\partial p_{j}}%
+o\left( \frac{1}{n}\right) ,
\end{equation*}%
where $J$ is the Fisher information matrix of the totality of probability
distributions supported on $\{1,\cdots ,d\}$. For $\left( J^{-1}\right)
^{i,j}=p_{i,\phi }\delta _{i,j}-p_{i,\phi }p_{j,\phi }$, we obtain the
lower-bound (\ref{bound}).

To prove the achievability, observe that the Schmidt coefficient is exactly
the spectrum of the reduced density matrix. As is discussed in \cite%
{KW,Matsumoto3},, the optimal measurement is the projectors $\left\{ 
\mathcal{W}_{\mathbf{n},A}\otimes \mathcal{W}_{\mathbf{n},B}\right\} $ ,
which is used in the protocol $\{C_{\ast }^{n}\}$, and the estimated of the
spectrum is $\frac{_{\mathbf{n}}}{n}$. It had been shown that the asymptotic
mean square error matrix of this estimate equals $\frac{1}{n}J^{-1}+o\left( 
\frac{1}{n}\right) $.\ Hence, if we estimate $\mathrm{H}(\mathbf{p}_{\phi })$
by $\mathrm{H}(\frac{_{\mathbf{n}}}{n})$, we can achieve the lower bound, as
is easily checked by using Taylor's expansion. Now, due to (\ref{grep-type}%
), our estimate $\frac{\log \dim \mathcal{V}_{\mathbf{n}}}{n}$ differs from $%
\mathrm{H}(\frac{_{\mathbf{n}}}{n})$ at most by $O\left( \frac{\log n}{n}%
\right) $. Therefore, their mean square error differs at most by $O\left( 
\frac{\log n}{n}\right) ^{2}=o\left( \frac{1}{n}\right) $. As a result, the
estimate based on $\{C_{\ast }^{n}\}$ achieves the lower-bound (\ref{bound}).
\end{pf}

\section{Conclusions and discussions}

\label{sec:conclusion}We have proposed a new protocol of entanglement
concentration $\{C_{\ast }^{n}\}$, which has the following properties.

\begin{enumerate}
\item The input state are many copies of unknown pure states.

\item The output is the exact maximally entangled state, and its Schmidt
rank.

\item Its performance is probabilistic, and entropy rate is asymptotically
achieved.

\item Any protocol is no better than a protocol given by modification of the
protocol $\{C_{\ast }^{n}\}$ in its classical output only.

\item The protocol is optimal up to higher orders or non-asymptotically,
depending on measures.

\item No classical communication is needed.

\item The classical output gives the estimate of the entropy of entanglement
with minimum asymptotic error, where minimum is taken over all the global
measurements.
\end{enumerate}

The key to the optimality arguments is Lemma~\ref{lem:k1}, which imply 4 in
above, and drastically simplified the arguments. As is pointed out
throughout the paper, almost all the statement of optimality, except for
Theorem~\ref{th:opt-exponent}, generalizes to the case where the Schmidt
coefficients are known.

As a measure of the distortion, we considered the worst-case distortion and
the average distortion. Trivially, the latter constraint is stronger, and
thus the proof for optimality was technically much simpler and results are
stronger. A problem is which one is more natural. This is very subtle
problem, but we think that the constraints on the average distortion is too
generous. The reason is as follows. Under this constraint, the strong
converse probability decreases very slowly (Theorem~\ref{opt-g-yield}, (ii)
). It is easy to generalize this statement to non-universal entanglement
concentration. This is in sharp contrast with the fact that strong converse
probability converges exponentially fast in many other information theoretic
problems.

Toward the end of Section~\ref{subsec:weighted}, using the linear
programming approach, we gave another proof of Theorem~\ref{opt-g-yield}.
The similar proof of Theorem~\ref{th:opt-exponent} is possible using the
Lagrangian%
\begin{equation*}
\mathfrak{L}^{\prime }\mathfrak{:}=\min_{U,V}\mathrm{E}_{Q_{C^{n}}^{U\otimes
V\phi }}^{X}f(X)-\lambda \max_{U,V}\left( \overline{\epsilon }%
_{C^{n}}^{U\otimes V\psi }-r_{n}\right) ,
\end{equation*}%
with $\mathrm{H}\left( \mathbf{p}_{\psi }\right) $ slightly smaller than $R$%
. In addition, using the Lagrangian%
\begin{eqnarray*}
&&\mathfrak{L}^{\prime \prime }\mathfrak{:}= \\
&&\min_{U,V}\mathrm{E}_{Q_{C^{n}}^{U\otimes V\phi }}^{X}\left\{ 
\begin{array}{c}
f(X) \\ 
+\lambda _{X}^{n}\left( r_{n}-\langle 2^{nX}\Vert \rho _{C^{n}}^{U\otimes
V\phi }(X)\Vert 2^{nX}\rangle \right)%
\end{array}%
\right\} ,
\end{eqnarray*}%
one can give another proof of the optimality results with the constraint on
worst-case distortion. It had been pointed out by many authors that the
theory of linear programming, especially the duality theorem, supplies
strong mathematical tool to obtain an upper/lower-bound. Our case is one of
such examples.

Almost parallel with the arguments in this paper, we can prove the
optimality of BBPS protocol for all the protocols which do not use
information about phases of Schmidt basis and Schmidt coefficients. For
that, we just have to replace average over all the local unitary in our
arguments with the one over the phases. This average kills all the coherence
between typical subspaces, changing the state to the direct sum of the
maximally entangled states, and we obtain an equivalence of our key lemma.
Rest of the arguments are also parallel, for an equivalence of (\ref%
{grep-type}) holds due to type theoretic arguments.

In the paper, we discussed universal entanglement concentration only, but
the importance of universal entanglement distillation is obvious. This topic
is already studied by some authors \cite{Brun, Pan}, but optimality of their
protocol, etc. are left for the future study.

Another possible future direction is to explore new applications of the
measurement used in our protocol. This measurement had already been applied
to the estimation of the spectrum~\cite{KW, Matsumoto3}, and the universal
data compression~\cite{Ha, JHHH}. In addition, after the appearance of the
first draft~\cite{HM} of this paper, the polynomial size circuit for this
measurement had been proposed~\cite{Aram}, meaning that this measurements
can be realized efficiently by forthcoming quantum computers.

\section*{Acknowledgment}

The main part of this research was conducted as a part of QCI project, JST,
led by Professor Imai. We are grateful for him for his supports. At that
time, MH was a member of a research group in RIKEN led by Professor S.
Amari, and MH is grateful for his generous support to this work. We are
thankful to Professor H. Nagaoka, Dr. T. Ogawa, Professor. M. Hamada for
discussions .K. M. is especially thankful to Prof. Nagaoka for his
implication that the average state over the unknown unitary may simplify the
optimality proof, and for his suggestion that the average yield may be
another natural measure.

\appendix

\section{Group representation theory}

\label{appendixA}

\begin{lem}
\label{lem:decohere} Let $U_{g}$ and $U_{g}^{\prime }$ be an irreducible
representation of $G$ on the finite-dimensional space $\mathcal{H}$ and $%
\mathcal{H}^{\prime }$, respectively. We further assume that $U_{g}$ and $%
U_{g}^{\prime }$ are not equivalent. If a linear operator $A$ in $\mathcal{H}%
\oplus \mathcal{H}^{\prime }$ is invariant by the transform $A\rightarrow
U_{g}\oplus U_{g}^{\prime }AU_{g}^{\ast }\oplus U_{g}^{^{\prime }\ast }$ for
any $g$, $\mathcal{H}A\mathcal{H^{\prime }}=0$. ~\cite{GW}
\end{lem}

\begin{lem}
\label{lem:shur} (Shur's lemma~\cite{GW}) Let $U_{g}$ be as defined in lemma~%
\ref{lem:decohere}. If a linear map $A$ in $\mathcal{H}$ is invariant by the
transform $A\rightarrow U_{g}AU_{g}^{\ast }$ for any $g$, $A=c\mathrm{Id}_{%
\mathcal{H}}$.
\end{lem}

\begin{lem}
\label{lem:shur2}Let $U_{g}$ be an irreducible representation of $G$ on the
finite dimensional space $\mathcal{H}$, and let $A$ be an linear map in $%
\mathcal{K}\otimes \mathcal{H}$. If $A$ is invariant by the transform $%
A\rightarrow I\otimes U_{g}AI\otimes U_{g}^{\ast }$ for any $g$, $A$ is the
form of $A^{\prime }\otimes \mathrm{Id}_{\mathcal{H}}$, with a linear map in 
$\mathcal{K}$.

\begin{pf}
Write $A=\sum_{i,j}A_{i}\otimes B_{j}$. Due to Shur's lemma, $B_{j}=c_{j}%
\mathrm{Id}_{\mathcal{H}}$. Therefore,%
\begin{equation*}
A=\sum_{i,j}A_{i}\otimes c_{j}\mathrm{Id}_{\mathcal{H}}=\left(
\sum_{i,j}c_{j}A_{i}\right) \otimes \mathrm{Id}_{\mathcal{H}},
\end{equation*}%
and we have the lemma.
\end{pf}
\end{lem}

\begin{lem}
\label{l5} If the representation $U_{g}($ $U_{h}^{\prime }$, resp.$)$ of $%
G(H $, resp.$)$ on the finite-dimensional space $\mathcal{H}(\mathcal{H}%
^{\prime }$, resp.$)$ is irreducible, the representation $U_{g}\times
U_{h}^{\prime }$ of the group $G\times H$ in the space $\mathcal{H}\otimes 
\mathcal{H}^{\prime }$ is also irreducible.
\end{lem}

\begin{pf}
Assume that if the representation $U_{g}\times U_{h}^{\prime }$ is
reducible, \textit{i. e.}, $\mathcal{H}\otimes \mathcal{H}^{\prime }$ has an
irreducible subspace $\mathcal{K}$. Denoting Haar measure in $G$ and $H$ by $%
\mu (\mathrm{d}g)$ and $\nu (\mathrm{d}h)$ respectively, Shur's lemma yields 
\begin{equation*}
\int U_{g}\otimes U_{h}^{\prime }|\phi \rangle \langle \phi |U_{g}^{\ast
}\otimes U_{h}^{^{\prime }\ast }\mu (\mathrm{d}g)\nu (\mathrm{d}h)=c\mathrm{%
Id}_{\mathcal{H}\otimes \mathcal{H}^{\prime }},
\end{equation*}%
for the RHS is invariant by both $U_{g}\,\cdot \,U_{g}^{\ast }$ and $%
U_{h}^{\prime }\,\cdot \,U_{h}^{^{\prime }\ast }$. This equation leads to $%
\int |\langle \psi |U_{g}\otimes U_{h}^{\prime }|\phi \rangle |^{2}\mu (%
\mathrm{d}g)\nu (\mathrm{d}h)=c$ and $c\dim \mathcal{H}\dim \mathcal{H}%
^{\prime }=\mu (G)\nu (H)\langle \phi |\phi \rangle $. Choosing $|\psi
\rangle $ from $\mathcal{K}^{\bot }$, the former equation gives $c=0$, which
contradicts with the latter.
\end{pf}

\section{\protect\bigskip Asymptotic yield of BBPS protocol}

\label{sec:cal-ave-ben}

From here to the end of the paper, we use the following notation.%
\begin{eqnarray*}
\mathbf{n}! &:&=\prod_{i}n_{i}!,\,\, \\
p_{i} &:&=p_{i}^{\phi },\,\mathbf{p}=(p_{1},\cdots ,p_{d})
\end{eqnarray*}%
In this section, we compute the yield of BBPS protocol, 
\begin{equation*}
\mathrm{E}_{\mathbf{p}}\left[ \frac{1}{n}\log \frac{n!}{\mathbf{n}!}\right]
=\sum_{\substack{ \mathbf{n:}  \\ \sum_{i}n_{i}=n,n_{i}\geq 0}}%
\prod\limits_{i}p_{i}^{n_{i}}\frac{n!}{\mathbf{n}!}\frac{1}{n}\log \left( 
\frac{n!}{\mathbf{n}!}\right)
\end{equation*}%
\label{bbps-yield}up to $O\left( \frac{1}{n}\right) $. Here, $\mathrm{E}_{%
\mathbf{p}}$ denotes average in terms of probability distribution 
\begin{equation*}
\mathbf{n}=(n_{1},\cdots ,n_{d})\sim \prod\limits_{i=1}^{d}p_{i}\,.
\end{equation*}%
Below, we assume $p_{i}\neq 0$. Due to Stirling's formula $n!=\sqrt{2\pi n}%
n^{n}e^{-n}\left( 1+O\left( \frac{1}{n}\right) \right) $,%
\begin{eqnarray*}
&&\frac{1}{n}\log \left( \frac{n!}{\mathbf{n}!}\right) \\
&=&\mathrm{H}\left( \frac{\mathbf{n}}{n}\right) -\frac{d-1}{2}\frac{\log n}{n%
} \\
&&-\frac{1}{n}\left( \frac{d-1}{2}\log 2\pi +\frac{1}{2}\sum%
\limits_{i=1}^{d}\log \frac{n_{i}}{n}\right) +R_{1}\left( \mathbf{n}\right) ,
\end{eqnarray*}%
where $R_{1}\left( \mathbf{n}\right) =\frac{1}{n}O\left( \max \left\{ \frac{1%
}{n},\frac{1}{n_{1}},\cdots ,\frac{1}{n_{d}}\right\} \right) $. Consider a
Taylor's expansion, 
\begin{eqnarray*}
&&\mathrm{H}\left( \frac{\mathbf{n}}{n}\right) \\
&=&\mathrm{H}\left( \mathbf{p}\right) +\sum_{i=1}^{d}\frac{\partial \mathrm{H%
}\left( \mathbf{p}\right) }{\partial p_{i}}\left( \frac{n_{i}}{n}%
-p_{i}\right) \\
&&+\frac{\log e}{2}\left( -\sum_{j=1}^{d}\frac{n_{j}^{2}}{p_{j}n^{2}}%
+1\right) +R_{2}\left( \frac{\mathbf{n}}{n},\mathbf{p}\right) ,
\end{eqnarray*}%
and let $R\left( \frac{\mathbf{n}}{n},\mathbf{p}\right) :=R_{1}(\mathbf{n}%
)+R_{2}(\frac{\mathbf{n}}{n},\mathbf{p})$. For $\mathrm{E}_{\mathbf{p}%
}f\left( \frac{n_{i}}{n}\right) =f(p_{i})+o(1)$ and $\mathrm{E}_{\mathbf{p}%
}n_{j}^{2}=n(n-1)p_{j}^{2}+np_{j}$, we have, 
\begin{eqnarray*}
&&\mathrm{E}_{\mathbf{p}}\left[ \frac{1}{n}\log \frac{n!}{\mathbf{n}!}\right]
\\
&=&\mathrm{H}\left( \mathbf{p}\right) -\frac{d-1}{2}\frac{\log n}{n} \\
&&-\frac{1}{n}\left\{ \frac{(d-1)}{2}\log 2\pi e+\frac{1}{2}%
\sum\limits_{i=1}^{d}\log p_{i}\right\} \\
&&+\mathrm{E}_{\mathbf{p}}R\left( \frac{\mathbf{n}}{n},\mathbf{p}\right)
+o\left( n^{-1}\right)
\end{eqnarray*}%
For 
\begin{equation}
\mathrm{E}_{\mathbf{p}}R(\frac{\mathbf{n}}{n},\mathbf{p})=o\left(
n^{-1}\right) ,  \label{remainder}
\end{equation}%
holds as is proved below, our calculation is complete.

For $\frac{1}{n}\log \frac{n!}{\mathbf{n}!}$ is bounded by constant, $%
R\left( \frac{\mathbf{n}}{n},\mathbf{p}\right) $ is bounded by a polynomial
function of $n$. Hence, due to the type theory, 
\begin{eqnarray*}
&&\mathrm{E}_{\mathbf{p}}R\left( \frac{\mathbf{n}}{n},\mathbf{p}\right) \\
&\leq &\mathrm{E}_{\mathbf{p}}\left[ \left. R\left( \frac{\mathbf{n}}{n},%
\mathbf{p}\right) \right\vert \left\Vert \frac{\mathbf{n}}{n}-\mathbf{p}%
\right\Vert <\delta \right] \\
&&+\mathrm{poly}\left( n\right) 2^{-nD\left( \delta \right) },
\end{eqnarray*}%
where $D\left( \delta \right) :=\min_{\mathbf{q\in }\left\{ \mathbf{q:}%
\left\Vert \mathbf{q}-\mathbf{p}\right\Vert <\delta \right\} }\mathrm{D}%
\left( \mathbf{q||p}\right) $. In the region $\left\{ \mathbf{n:}\left\Vert 
\frac{\mathbf{n}}{n}-\mathbf{p}\right\Vert <\delta \right\} ,$%
\begin{eqnarray*}
\left\vert R_{1}(\mathbf{n)}\right\vert &=&O\left( \frac{1}{n^{2}}\right) ,
\\
\left\vert R_{2}\left( \frac{\mathbf{n}}{n},\mathbf{p}\right) \right\vert
&\leq &\sum_{i,j,k}\max_{\mathbf{p}_{0}:\left\Vert \mathbf{p}_{0}-\mathbf{p}%
\right\Vert <\delta }\left\vert \frac{\partial ^{3}H\left( \mathbf{p}%
_{0}\right) }{\partial p_{i}\partial p_{j}\partial p_{k}}\right\vert \delta
^{3},
\end{eqnarray*}%
Observe also $D\left( \delta \right) =O\left( \delta ^{2}\right) .$ Hence,
if $\delta =n^{-\frac{3}{8}}$, $\left\vert R_{2}\left( \frac{\mathbf{n}}{n},%
\mathbf{p}\right) \right\vert =o\left( n^{-1}\right) $ in the region $%
\left\{ \mathbf{n:}\left\Vert \frac{\mathbf{n}}{n}-\mathbf{p}\right\Vert
<\delta \right\} $, and $2^{-nD\left( \delta \right) }=2^{-O\left(
n^{2/8}\right) }$, implying (\ref{remainder}).

\section{Difference between the average yield of BBPS protocol and $\left\{
C_{\ast }^{n}\right\} $}

\label{sec:diff-ave-yield}

$\dim \mathcal{V}_{\mathbf{n}}$ and $a_{\mathbf{n}}^{\phi }$ are explicitly
given as follows.

\begin{eqnarray}
\dim \mathcal{V}_{\mathbf{n}} &=&\sum_{\pi \in S_{n}}\mathrm{sgn}(\pi )\frac{%
n!}{(\mathbf{n}+\mathbf{\delta }-\pi (\mathbf{\delta }))!},  \label{dimV-1}
\\
&=&\frac{\left( n+\frac{d(d-1)}{2}\right) !}{(\mathbf{n}+\mathbf{\delta )}!}%
\prod\limits_{i,j:i>j}(n_{i}-n_{j}+j-i)  \label{dimV-2} \\
a_{\mathbf{n}}^{\phi } &=&\frac{\dim \mathcal{V}_{\mathbf{n}}}{%
\prod\limits_{i,j:i>j}(p_{i}-p_{j})}\sum_{\pi \in S_{n}}\mathrm{sgn}(\pi
)\prod\limits_{i}p_{i}^{n_{\pi (i)}+\delta _{\pi (i)}},  \notag
\end{eqnarray}%
where

\begin{eqnarray*}
\mathbf{\delta } &:&=(d-1,d-2,\cdots ,0), \\
\pi (\mathbf{\delta }) &:&=\left( \delta _{\pi (1)},\delta _{\pi (1)},\cdots
,\delta _{\pi (d)}\right) .
\end{eqnarray*}%
Below, $p_{i}^{\phi }\neq p_{j}^{\phi }$, and $p_{i}^{\phi }\neq 0$ are
assumed for simplicity. The average yield equals

\begin{eqnarray*}
&&\frac{1}{n}\sum_{\mathbf{n}}a_{\mathbf{n}}^{\phi }\log (\dim \mathcal{V}_{%
\mathbf{n}}) \\
&=&\frac{1}{n}\frac{1}{\prod\limits_{i,j:i>j}(p_{i}-p_{j})}\sum_{\pi _{0}\in
S_{n}}\mathrm{sgn}(\pi _{0}) \\
&&\times \sum_{\mathbf{n}}\prod\limits_{i}p_{i}^{n_{\pi _{0}(i)}+\delta
_{\pi _{0}(i)}}(\dim \mathcal{V}_{\mathbf{n}})\log (\dim \mathcal{V}_{%
\mathbf{n}}),
\end{eqnarray*}%
where $\mathbf{n}$ is summed over the region satisfying (\ref{yungindex}).
In the sum over $\pi _{0}$, we first compute the term for $\pi _{0}=\mathrm{%
id}$ (other terms will turn out to be exponentially small):

\begin{eqnarray}
&&\sum_{\mathbf{n}}\prod\limits_{i}p_{i}^{n_{i}+\delta _{i}}(\dim \mathcal{V}%
_{\mathbf{n}})\log (\dim \mathcal{V}_{\mathbf{n}})  \notag \\
&=&\sum_{\pi \in S_{n}}\mathrm{sgn}(\pi )\prod\limits_{i}p_{i}^{\delta _{\pi
(i)}}\sum_{\mathbf{n}}\prod\limits_{i}p_{i}^{n_{i}+\mathbf{\delta }%
_{i}-\delta _{\pi (i)}}  \notag \\
&&\times \frac{n!}{(\mathbf{n}+\mathbf{\delta }-\pi (\mathbf{\delta }))!}%
\log \left\{ \sum_{\pi ^{\prime }\in S_{n}}\mathrm{sgn}(\pi ^{\prime })\frac{%
n!}{(\mathbf{n}+\mathbf{\delta }-\pi ^{\prime }(\mathbf{\delta }))!}\right\}
\notag \\
&=&\sum_{\pi \in S_{n}}\mathrm{sgn}(\pi )\prod\limits_{i}p_{i}^{\delta _{\pi
(i)}}\sum_{\mathbf{n}^{\pi }}\prod\limits_{i}p_{i}^{n_{i}^{\pi }}\frac{n!}{%
\mathbf{n}^{\pi }!}  \notag \\
&&\times \log \left\{ \sum_{\pi ^{\prime }\in S_{n}}\mathrm{sgn}(\pi
^{\prime })\frac{n!}{(\mathbf{n}^{\pi }+\pi (\mathbf{\delta )}-\pi ^{\prime
}(\mathbf{\delta }))!}\right\} ,  \label{ave-y-cal}
\end{eqnarray}%
where $\mathbf{n}^{\pi }$ is defined by $\mathbf{n}^{\pi }:=\mathbf{n}+%
\mathbf{\delta }-\pi (\mathbf{\delta })$. For the probability sharply
concentrates at the neighborhood of $\frac{\mathbf{n}^{\pi }}{n}=\mathbf{p}$%
, we have, 
\begin{equation}
\sum_{\mathbf{n}^{\pi }}\prod\limits_{i}p_{i}^{n_{i}^{\pi }}\frac{n!}{%
\mathbf{n}^{\pi }!}f\left( \frac{\mathbf{n}^{\pi }}{n}\right) =f(\mathbf{p}%
)+O(2^{-cn}).  \label{lawLN}
\end{equation}%
The main part of (\ref{ave-y-cal}) rewrites,%
\begin{eqnarray*}
&&\sum_{\mathbf{n}^{\pi }}\prod\limits_{i}p_{i}^{n_{i}^{\pi }}\frac{n!}{%
\mathbf{n}^{\pi }!}\log \left\{ \sum_{\pi ^{\prime }\in S_{n}}\mathrm{sgn}%
(\pi ^{\prime })\frac{n!}{(\mathbf{n}^{\pi }+\pi (\mathbf{\delta )}-\pi
^{\prime }(\mathbf{\delta }))!}\right\} \\
&=&\sum_{\mathbf{n}^{\pi }}\prod\limits_{i}p_{i}^{n_{i}^{\pi }}\frac{n!}{%
\mathbf{n}^{\pi }!}\left\{ 
\begin{array}{c}
\log \frac{n!}{\mathbf{n}^{\pi }!} \\ 
+\log \sum\limits_{\pi ^{\prime }\in S_{n}}\mathrm{sgn}(\pi ^{\prime })\frac{%
\mathbf{n}^{\prime }!}{(\mathbf{n}^{\pi }+\pi (\mathbf{\delta )}-\pi
^{\prime }(\mathbf{\delta }))!}%
\end{array}%
\right\} .
\end{eqnarray*}%
The first term is exponentially close to $n$ times the average yield of BBPS
protocol. The second term is, 
\begin{eqnarray*}
&&\sum_{\mathbf{n}^{\pi }}\prod\limits_{i}p_{i}^{n_{i}^{\pi }}\frac{n!}{%
\mathbf{n}^{\pi }!}\log \sum_{\pi ^{\prime }\in S_{n},}\mathrm{sgn}(\pi
^{\prime })\frac{\mathbf{n}^{\pi }!}{(\mathbf{n}^{\pi }+\pi (\mathbf{\delta )%
}-\pi ^{\prime }(\mathbf{\delta }))!} \\
&=&\sum_{\mathbf{n}^{\pi }}\prod\limits_{i}p_{i}^{n_{i}^{\pi }}\frac{n!}{%
\mathbf{n}^{\pi }!}\times \\
&&\log \sum_{\pi ^{\prime }\in S_{n}}\mathrm{sgn}(\pi ^{\prime })\frac{%
\prod\limits_{i:\delta _{\pi (i)}-\delta _{\pi ^{\prime
}(i)}<0}\prod\limits_{j=1}^{\delta _{\pi (i)}-\delta _{\pi ^{\prime
}(i)}}(n_{i}^{\pi }-j+1)}{\prod\limits_{i:\delta _{\pi (i)}-\delta _{\pi
^{\prime }(i)}>0}\prod\limits_{j=1}^{\delta _{\pi ^{\prime }(i)}-\delta
_{\pi (i)}}(n_{i}^{\pi }+j)} \\
&=&\log \sum_{\pi ^{\prime }\in S_{n}}\mathrm{sgn}(\pi ^{\prime })\frac{%
\prod\limits_{i:\delta _{\pi (i)}-\delta _{\pi ^{\prime
}(i)}<0}p_{i}^{\delta _{\pi ^{\prime }(i)}-\delta _{\pi (i)}}}{%
\prod\limits_{i:\delta _{\pi (i)}-\delta _{\pi ^{\prime
}(i)}>0}p_{i}^{\delta _{\pi (i)}-\delta _{\pi ^{\prime }(i)}}}+o(1) \\
&=&\log \sum_{\pi ^{\prime }\in S_{n}}\mathrm{sgn}(\pi ^{\prime
})\prod_{i}p_{i}^{\delta _{\pi ^{\prime }(i)}-\delta _{\pi (i)}}+o(1) \\
&=&\log \prod\limits_{i,j:i>j}(p_{i}-p_{j})\prod_{k}p_{k}^{-\delta _{\pi
(k)}}+o(1),
\end{eqnarray*}%
where the second equation is due to (\ref{lawLN}).

To sum up, the term for $\pi _{0}=\mathrm{id}$ is

\begin{eqnarray*}
&&\frac{1}{\prod\limits_{i,j:i>j}(p_{i}-p_{j})}\sum_{\pi \in S_{n}}\mathrm{%
sgn}(\pi )\prod\limits_{i}p_{i}^{\delta _{\pi (i)}} \\
&&\times \left\{ 
\begin{array}{c}
\frac{1}{n}\left\{ 
\begin{array}{c}
\log \left\{ \prod\limits_{i,j:i>j}(p_{i}-p_{j})\prod_{k}p_{k}^{-\delta
_{\pi (k)}}\right\} \\ 
+o\left( 1\right)%
\end{array}%
\right\} \\ 
+\mbox{average yield of BBPS}%
\end{array}%
\right\} \\
&=&\frac{1}{n\prod\limits_{i,j:i>j}(p_{i}-p_{j})}\sum_{\pi \in S_{n}}\mathrm{%
sgn}(\pi )\prod\limits_{i}p_{i}^{\delta _{\pi (i)}} \\
&&\times \log \left\{
\prod\limits_{i,j:i>j}(p_{i}-p_{j})\prod_{k}p_{k}^{-\delta _{\pi
(k)}}\right\} \\
&&+\mbox{average yield of BBPS}+o\left( \frac{1}{n}\right)
\end{eqnarray*}

The terms for $\pi _{0}\neq $\textrm{id }are of the form 
\begin{eqnarray*}
&&\sum_{\mathbf{n}^{\pi }}\prod\limits_{i}p_{i}^{n_{\pi _{0}(i)}^{\pi }}%
\frac{n!}{\mathbf{n}^{\pi }!}f\left( \frac{\mathbf{n}^{\pi }}{n}\right) \\
&=&\sum_{\mathbf{n}^{\pi }}\prod\limits_{i}p_{\pi _{0}^{-1}(i)}^{n_{i}^{\pi
}}\frac{n!}{\mathbf{n}^{\pi }!}f\left( \frac{\mathbf{n}^{\pi }}{n}\right) .
\end{eqnarray*}%
Observe that the probability distribution $\prod\limits_{i}p_{\pi
_{0}(i)}^{n_{i}^{\pi }}\frac{n!}{\mathbf{n}^{\pi }!}$ is concentrated around 
$\mathbf{n}^{\pi }=n\pi _{0}^{-1}(\mathbf{p})$, which is not close to the
region where $\mathbf{n}^{\pi }$ takes its value. Hence, for $f(x)$ is
bounded by a constant, due to the large deviation principles, this sum
should be exponentially small.

\section{Asymptotic average yield of the optimal non-universal concentration}

\label{sec:cal-ave-opt} In this section, we discuss the asymptotic
performance of a optimal non-universal entanglement concentration, or an
optimal entanglement concentration for known input state. In terms of error
probability, intensive research is done by \cite{Morikoshi}. Here, our
concern is average yield of the optimal protocol.

For that, we use Hardy's formula\thinspace \cite{Hardy} : the average of
number of bell pairs concentrated from known pure state is.

\begin{equation}
\sum\limits_{i=0}^{m}\left( \alpha _{i}-\alpha _{i+1}\right) T_{i}\log T_{i},
\label{hardy}
\end{equation}%
where $\alpha _{i}$ is a Schmidt coefficient (in decreasing order, with the
convention $\alpha _{m+1}=0$), and $T_{i}$ is the number of Schmidt basis
such that its corresponding Schmidt coefficient is larger than or equal to $%
\alpha _{i}$.

In this appendix, we evaluate (\ref{hardy}) asymptotically for qubit
systems, assuming that an input state is $\left\vert \phi \right\rangle =%
\sqrt{q}\left\vert 0\right\rangle +\sqrt{p}\left\vert 1\right\rangle $ with $%
p>q$. Below, we always neglect $o(1)$ terms, unless otherwise mentioned,
because this quantity is the yield multiplied by $n$.%
\begin{align*}
& \sum_{i=0}^{n-1}\left( p^{i}q^{n-i}-p^{i+1}q^{n-i-1}\right)
\sum\limits_{j=0}^{i}\binom{n}{j}\log \sum\limits_{k=0}^{i}\binom{n}{k} \\
& =\left( 1-\frac{p}{q}\right)
\sum_{i=0}^{n-1}\sum\limits_{j=0}^{i}p^{i}q^{n-i}\binom{n}{j}\log
\sum\limits_{k=0}^{i}\binom{n}{k} \\
& =\left( 1-\frac{p}{q}\right)
\sum_{j=0}^{n-1}\sum\limits_{i=j}^{n}p^{i}q^{n-i}\binom{n}{j}\log
\sum\limits_{k=0}^{i}\binom{n}{k} \\
& =\left( 1-\frac{p}{q}\right) \sum_{j=0}^{n-1}p^{j}q^{n-j}\binom{n}{j}%
\sum\limits_{l=0}^{n-j}\left( \frac{p}{q}\right) ^{l}\log
\sum\limits_{k=0}^{j+l}\binom{n}{k}
\end{align*}

For $\sum\limits_{l=0}^{n-j}\left( \frac{p}{q}\right) ^{l}\log
\sum\limits_{k=0}^{j+l}\binom{n}{k}$ is at most polynomial order, due to the
large deviation principles, the range of $j$ can be replaced by $[n(p-\delta
),\,n(p+\delta )]$. In this region,. $j+l$ is the order of $n$. Also, the
range of $l$ can be replaced by $[1,\,n^{1/2})$, for 
\begin{align*}
& \sum\limits_{l=n/2}^{n-j}\left( \frac{p}{q}\right) ^{l}\log
\sum\limits_{k=0}^{j+l}\binom{n}{k} \\
& \leq \left( \frac{p}{q}\right) ^{n^{1/2}}\times n\times \mathrm{poly}%
(n)=o(1).
\end{align*}%
Hence, we evaluate $\frac{\sum\limits_{k=0}^{np^{\prime }}\binom{n}{k}}{%
\binom{n}{np^{\prime }}}$ with $0<p^{\prime }<\frac{1}{2}$. First, we
upper-bound $\frac{\sum\limits_{k=0}^{np^{\prime }-n^{1/3}}\binom{n}{k}}{%
\binom{n}{np^{\prime }}}$ by 
\begin{equation*}
\frac{\sum\limits_{k=0}^{np^{\prime }-n^{1/3}}2^{nh(\frac{k}{n})}}{\frac{1}{%
n+1}2^{nh(p^{\prime })}}\leq \frac{(n+1)2^{nh(\frac{np^{\prime }-n^{1/3}}{n}%
)}}{\frac{1}{n+1}2^{nh(p^{\prime })}}=(n+1)^{2}2^{-O\left( n^{1/3}\right) },
\end{equation*}%
meaning this part is negligible. Hence, we have 
\begin{eqnarray*}
&&\frac{\sum\limits_{k=0}^{np^{\prime }}\binom{n}{k}}{\binom{n}{np^{\prime }}%
}=\frac{\sum\limits_{k=0}^{n^{1/3}}\binom{n}{np^{\prime }-k}}{\binom{n}{%
np^{\prime }}} \\
&=&\sum\limits_{k=0}^{n^{1/3}}\frac{\left( np^{\prime }\right) !\left(
nq^{\prime }\right) !}{\left( np^{\prime }-k\right) !\left( nq^{\prime
}+k\right) !} \\
&=&1+\sum\limits_{k=1}^{n^{1/3}}\prod_{i=1}^{k}\frac{np^{\prime }-k+i}{%
nq^{\prime }+i} \\
&=&\sum\limits_{k=0}^{n^{1/3}}\left( \frac{p^{\prime }}{q^{\prime }}\right)
^{k}\left( 1+\frac{O(n^{1/3}}{n})\,\right) ^{n^{1/3}}=\frac{q^{\prime }}{%
q^{\prime }-p^{\prime }}\;\left( 1+o(1)\right) .
\end{eqnarray*}

Hence, the average yield is, 
\begin{equation}
\left( 1-\frac{p}{q}\right) \sum\limits_{l=0}^{n^{1/2}}\left( \frac{p}{q}%
\right) ^{l}\sum_{j}p^{j}q^{n-j}\binom{n}{j}\left\{ 
\begin{array}{c}
\log \binom{n}{j+l} \\ 
+\log \frac{1-\left( j+l\right) /n}{1-2\left( j+l\right) /n}%
\end{array}%
\right\} .  \label{ave-opt-2}
\end{equation}%
The second term of this is evaluated by using the following identity,%
\begin{equation}
\left( 1-\frac{p}{q}\right) \sum\limits_{l=0}^{n^{1/2}}\left( \frac{p}{q}%
\right) ^{l}f\left( x+l/n\right) =f(x)+o(1),  \label{sum-p/q}
\end{equation}%
where $f$ is continuous and bounded by a polynomial function. This identity
holds true because of the upper-bound to the RHS,%
\begin{align*}
& \left( 1-\frac{p}{q}\right) \sum\limits_{l=0}^{n^{1/2}}\left( \frac{p}{q}%
\right) ^{l}\max_{y:y\in \lbrack 0,n^{-1/2}]}f\left( x+y\right) \\
& \leq \max_{y:y\in \lbrack 0,n^{-1/2}]}f\left( x+y\right) ,
\end{align*}%
and the lower-bound the RHS, 
\begin{align*}
& \left( 1-\frac{p}{q}\right) \sum\limits_{l=0}^{n^{1/2}}\left( \frac{p}{q}%
\right) ^{l}\max_{y:y\in \lbrack 0,n^{-1/2}]}f\left( x+y\right) \\
& =\left( 1-\left( \frac{p}{q}\right) ^{n^{1/2}+1}\right) \max_{y:y\in
\lbrack 0,n^{-1/2}]}f\left( x+y\right) .
\end{align*}%
Hence, the second term of (\ref{ave-opt-2}), or 
\begin{equation*}
\left( 1-\frac{p}{q}\right) \sum\limits_{l=0}^{n^{1/2}}\left( \frac{p}{q}%
\right) ^{l}\log \frac{q-l/n}{1-2(p+l/n)}
\end{equation*}%
equals, due to Eq. (\ref{sum-p/q}), $\log \frac{q}{q-p}+o(1).$

The first term of (\ref{ave-opt-2}) is evaluated as follows. Due to
Stirling's formula and Taylor's expansion,%
\begin{align*}
& \log \binom{n}{j+l} \\
& =n\mathrm{h}(p)-\left( \log p+\log e\right) \left( j+l-np\right) \\
& -\left( \log q+\log e\right) \left( n-j-l-nq\right) \\
& +\frac{\log e}{2}\left( -\frac{(j+l)^{2}}{pn}-\frac{(n-j-l)^{2}}{qn}%
+n\right) +nR_{2}(\frac{\mathbf{n}}{n},\mathbf{p}) \\
& -\frac{\log n}{2} \\
& -\frac{1}{2}\left( \log 2\pi +\log \frac{j+l}{n}+\log \left( 1-\frac{j+l}{n%
}\right) \right) +nR_{1}\left( \mathbf{n}\right) ,
\end{align*}%
whose average by the binomial distribution $p^{j}q^{n-j}\binom{n}{j}$ is,%
\begin{align*}
& n\mathrm{h}(p)-(\log p-\log q)l \\
& +\frac{\log e}{2}\left( -\frac{1}{n}\frac{l^{2}}{pq}\right) -\frac{\log n}{%
2} \\
& -\frac{1}{2}\left( \log 2\pi e+\log \left( p+\frac{l}{n}\right) +\log
\left( q-\frac{l}{n}\right) \right) \\
& +nR_{1}(\mathbf{n})+nR_{2}(\frac{\mathbf{n}}{n},\mathbf{p}).
\end{align*}%
Due to Eq. (\ref{sum-p/q}), multiplied by $(p/q)^{l}$ and summed over $l$,
the first term is obtained as: 
\begin{align*}
& n\mathrm{h}(p)-\frac{\log n}{2} \\
& -\frac{1}{2}\left( \log 2\pi e+\log p+\log q\right) \\
& -(\log p-\log q)\frac{p/q}{1-p/q} \\
& +nR_{1}(\mathbf{n})+nR_{2}\left( \frac{\mathbf{n}}{n},\mathbf{p}\right)
\end{align*}%
We can prove $nR_{1}(\mathbf{n})+nR_{2}\left( \frac{\mathbf{n}}{n},\mathbf{p}%
\right) $ is negligible almost in the same way as in Appendix \ref%
{sec:cal-ave-ben}. After all, the average yield is, 
\begin{align*}
& \mathrm{h}(p)-\frac{\log n}{2n}+ \\
& \frac{1}{n}\left\{ 
\begin{array}{c}
-\frac{1}{2}\log 2\pi eqp \\ 
+\frac{p/q}{1-p/q}\log \frac{q}{p}+\log \frac{1}{1-p/q}%
\end{array}%
\right\} +o\left( \frac{1}{n}\right) .
\end{align*}

\section{$\varlimsup_{n\rightarrow \infty }\frac{1}{n}\mathrm{D}(Q_{C_{\ast
}^{n}}^{\protect\psi }\Vert Q_{C_{\ast }^{n}}^{\protect\phi })$}

\label{sec:cal-divergence}Let us define,%
\begin{eqnarray*}
\mathbf{p} &:&=\mathbf{p}^{\phi },\,\mathbf{q}:=\mathbf{p}^{\psi },\mathbf{l}%
:=\mathbf{n}+\mathbf{\delta ,} \\
f\left( \mathbf{l}\right) &:&=\log \frac{\prod\limits_{i,j:i>j}(p_{i}-p_{j})%
\sum\limits_{\pi \in S_{n}}\mathrm{sgn}(\pi )\prod\limits_{i}q_{i}^{l_{\pi
(i)}}}{\prod\limits_{i,j:i>j}(q_{i}-q_{j})\sum\limits_{\pi \in S_{n}}\mathrm{%
sgn}(\pi )\prod\limits_{i}p_{i}^{l_{\pi (i)}}}
\end{eqnarray*}%
and our task is to compute, 
\begin{eqnarray*}
&&\mathrm{D}(Q_{C_{\ast }^{n}}^{\psi }\Vert Q_{C_{\ast }^{n}}^{\phi }) \\
&=&\sum_{\mathbf{n}}\frac{\dim \mathcal{V}_{\mathbf{n}}}{\prod%
\limits_{i,j:i>j}(q_{i}-q_{j})}\sum_{\pi \in S_{n}}\mathrm{sgn}(\pi
)\prod\limits_{i}q_{i}^{l_{\pi (i)}}f\left( \mathbf{l}\right) .
\end{eqnarray*}%
Due to the argument stated at the end of Appendix \ref{sec:diff-ave-yield},
in the first sum over $\pi \in S_{n}$, we can concentrate on the term $\pi =%
\mathrm{id}$, which is evaluated as follows. 
\begin{eqnarray*}
&&\sum_{\mathbf{n}}\frac{\dim \mathcal{V}_{\mathbf{n}}}{\prod%
\limits_{i,j:i>j}(q_{i}-q_{j})}\prod\limits_{i}q_{i}^{l_{i}}f\left( \mathbf{l%
}\right) \\
&=&\sum_{\mathbf{l}}\frac{1}{\prod\limits_{i,j:i>j}(q_{i}-q_{j})}\frac{%
\left( n+\frac{d(d-1)}{2}\right) !}{\mathbf{l}!} \\
&&\times \frac{\prod\limits_{i,j:i>j}(l_{i}-l_{j})}{\prod\limits_{i:i=0}^{%
\frac{d(d-1)}{2}}\left( n+\frac{d(d-1)}{2}-i\right) }\prod%
\limits_{i}q_{i}^{l_{i}}f\left( \mathbf{l}\right) \\
&=&\frac{\prod\limits_{i,j:i>j}(q_{i}-q_{j})}{\prod%
\limits_{i,j:i>j}(q_{i}-q_{j})}f\left( \left( n+\frac{d\left( d-1\right) }{2}%
\right) \mathbf{q}\right) +o\left( f\left( n\right) \right) \\
&=&f\left( \left( n+\frac{d\left( d-1\right) }{2}\right) \mathbf{q}\right)
+o\left( f\left( n\right) \right)
\end{eqnarray*}%
Here, the second equation was derived as follows. We first extended the
region of $\mathbf{l}$ to $\left\{ \mathbf{l};\sum_{i}l_{i}=n+\frac{d\left(
d-1\right) }{2}\right\} $, because this causes only exponentially small
difference due to the large deviation principles. Then, we applied the law
of the large number. Observe that 
\begin{eqnarray*}
f\left( \mathbf{l}\right) &=&\log \frac{\prod\limits_{i,j:i>j}(p_{i}-p_{j})}{%
\prod\limits_{i,j:i>j}(q_{i}-q_{j})}+\log \frac{\prod\limits_{i}q_{i}^{l_{i}}%
}{\prod\limits_{i}p_{i}^{l_{i}}} \\
&&+\log \frac{1+\sum\limits_{\substack{ \pi \in S_{n}  \\ \pi \neq \mathrm{id%
} }}\mathrm{sgn}(\pi )\prod\limits_{i}q_{i}^{l_{\pi (i)}-l_{i}}}{%
1+\sum\limits _{\substack{ \pi \in S_{n}  \\ \pi \neq \mathrm{id}}}\mathrm{%
sgn}(\pi )\prod\limits_{i}p_{i}^{l_{\pi (i)}-l_{i}}} \\
&=&\log \frac{\prod\limits_{i}q_{i}^{l_{i}}}{\prod\limits_{i}p_{i}^{l_{i}}}%
+O(1),
\end{eqnarray*}%
where the second equation is true due to the inequality, 
\begin{equation*}
1\leq 1+\sum\limits_{\substack{ \pi \in S_{n},  \\ \pi \neq \mathrm{id}}}%
\mathrm{sgn}(\pi )\prod\limits_{i}q_{i}^{l_{\pi (i)}-l_{i}}\leq 1+d!.
\end{equation*}%
After all, we have,%
\begin{eqnarray*}
&&\mathrm{D}(Q_{C_{\ast }^{n}}^{\psi }\Vert Q_{C_{\ast }^{n}}^{\phi }) \\
&=&\left( n+\frac{d(d-1)}{2}\right) \log \frac{\prod\limits_{i}q_{i}^{q_{i}}%
}{\prod\limits_{i}p_{i}^{p_{i}}}+O(1) \\
&=&n\mathrm{D}(\mathbf{q}\Vert \mathbf{p})+O(1).
\end{eqnarray*}

\end{document}